# Compression-Driven Anomaly Detection in Brain MRI Using an Interpretable Quantum Autoencoder

Santanu Ganguly[0000-0003-0141-9228]1, Xing Liang*[0000-0002-6630-298X]and Dimitrios Makris[0000-0001-6170-0236]

Kingston University, Kingston upon Thames KT1 2EE, UK
Quantum AI (QAI) Research Group
School of Computer Science and Mathematics (CSM)
Faculty of Engineering, Computing and the Environment (ECE)
[1]ganguly.santanu@kingston.ac.uk
*x.liang@kingston.ac.uk

**Abstract.** We study a quantum autoencoder (QAE) for compression-driven anomaly detection in brain MRI data. The approach leverages angle encoding to map image patches into quantum states, followed by a variational encoder–decoder architecture trained to discard information via auxiliary trash qubits. Anomaly scores reflect the degree to which inputs resist compression relative to normal data, with higher scores corresponding to deviations from the learned normal manifold. Evaluated on publicly available brain MRI DICOM datasets, the method achieves a slice-level ROC-AUC of ~0.95 and a patch-level ROC-AUC of ~0.813, outperforming classical autoencoder and PCA baselines. Analysis of the learned parameters reveals a pronounced encoder–decoder asymmetry, where effective anomaly detection arises from structured information compression within the encoder rather than increased parameter magnitude or decoder expressivity. This results in a controlled compression–reconstruction trade-off with a clear operating regime that supports principled threshold selection. Qualitative evaluation further shows that the QAE produces spatially localized anomaly heatmaps aligned with tumorous regions. The results, supported by promising baseline performances, demonstrate that quantum autoencoders provide an interpretable and controllable mechanism for anomaly detection based on incompressibility with respect to a learned latent representation. This work highlights the potential of quantum autoencoders as a principled tool for studying compression dynamics in quantum machine learning, with promising implications for decision-support in medical imaging workflows.

## 1 Introduction

Quantum machine learning (QML) is an interdisciplinary field that unites quantum computing with classical machine learning, two of the most dynamic areas in contemporary research. Advances in quantum computation and quantum information theory have already influenced multiple domains of computer science, including information theory [1], search functions [2, 31], emotion representation, and image processing [6, 17-30]. By exploiting quantum mechanical principles such as superposition, entanglement and quantum interference, quantum computers offer new computational paradigms capable of addressing certain classes of problems that are inefficient or intractable on classical architectures [1-7].

Within this broader context, QML aims to integrate quantum computational models with classical learning frameworks in order to develop algorithms capable of processing and learning from complex data structures. One prominent class of models in the area of QML is the quantum neural network (QNN), which is commonly defined as a parameterized quantum circuit with trainable continuous parameters [16-22, 60-62]. These models are designed to address classically challenging problems that demand exponential memory or representational capacity, positioning QNNs as a natural evolution of neurocomputing systems [16].

Recent developments on quantum computing [3, 87] have explored quantum-enhanced variants of well-established machine learning paradigms, including neural networks, autoencoders, reinforcement learning, adversarial learning, and transformers [48, 50, 64, 88-91]. In most current implementations, hybrid classical-quantum training frameworks are employed, where quantum circuits perform data transformations while classical optimization algorithms update circuit parameters. In the domain of medical imaging, studies on QAE remain rare to date, with Singh et al. [71] reporting on MedMNIST [55, 56] datasets, and comparing with their classical counterparts of Resnet-18 [114] and Resnet-50 [114].

### 1.1 Quantum Autoencoders: Motivation

Recent studies in quantum machine learning (QML) suggest that parameterized quantum circuits, optimized with classical routines and related frameworks such as quantum circuit learning, can support learning in high-dimensional feature spaces while using substantially fewer parameters than many classical models [92–95]. Among these approaches, the quantum autoencoder (QAE) is particularly attractive because it compresses data into a reduced number of qubits and reconstructs it using a trained quantum circuit, with classical computation used mainly for parameter optimization [48]. Existing QAE research has focused largely on data compression, feature transformation, and anomaly detection in simplified or single-class grayscale image settings [72, 79, 100–104], while broader theoretical work suggests that quantum-enhanced variants of established machine-learning paradigms may offer advantages as quantum hardware matures [3, 48, 50, 64, 87–91].

Whereas classical deep learning models have been used to study brain MRI data [51, 124, 134, 135], they often require large parameter counts and substantial computational resources, and lack interpretability. In contrast, quantum autoencoders models provide an alternative route to compact representation learning, but most existing studies address either compression or classification in isolation and rely on simplified benchmarks, or heavily downscaled inputs. This work is motivated by the quest to test whether hybrid classical–quantum architectures can learn meaningful representations from realistic medical imaging data under controlled conditions. To this end, we apply a Romero *et al*. [48] type QAE-based model to high-quality grayscale Brain MRI DICOM medical images, and conduct anomaly detection through controlled information compression rather than explicit reconstruction optimization.

### 1.2 Contributions

The contributions of this work lie in the use of a **quantum autoencoder as a compression-driven anomaly detector for brain MRI data**, with emphasis on mechanism, interpretability, and spatial anomaly localization rather than on reconstruction fidelity alone. More specifically, the contribution is novel in four respects:

I. **Empirical contribution: QAE for DICOM data.** To the best of our knowledge, this study is among the first to apply a quantum autoencoder (QAE) to brain MRI anomaly detection using DICOM-based data. The method is designed for detecting deviations between normal and tumorous brain images in an unsupervised setting. We evaluate a general QAE framework following Romero *et al*. [48] for anomaly detection on high-resolution grayscale DICOM brain MRI data and compare its performance with a classical autoencoder, reporting encouraging results with a maximum slice-ROC of ~0.95 and a patch-level ROC-AUC of 81.3.

II. **Methodological contribution: Compression-based anomaly scoring via trash qubits.** Unlike classical autoencoder approaches that rely primarily on reconstruction error, the proposed method uses explicit information compression by driving auxiliary ("trash") qubits toward a fixed reference state. Anomaly scores therefore reflect incompressibility with respect to a learned latent representation, providing a distinct detection mechanism.

III. **Interpretability contribution: Mechanistic analysis of learned quantum parameters.** A central novelty is not only the application itself, but also the interpretation of how the QAE works. The study shows that anomaly detection performance is linked to a structured quantum encoder–decoder asymmetry, where compression emerges through the layer-wise organization of encoder parameters at a quantum level rather than through increased overall parameter magnitude or decoder expressivity. This offers a *circuit-level explanation* of the compression–reconstruction trade-off.

IV. **Application contribution: Spatially resolved anomaly heatmaps from a QAE pipeline.** The framework produces localized anomaly heatmaps on MRI slices, allowing visual identification of tumorous regions. This extends the QAE from a purely abstract latent-space model to a practically interpretable tool for medical-image analysis.

### 1.3 Paper Organization

This paper is organized as follows: section 2 reviews related work on classical and quantum autoencoders (QAEs), outlining their architectures and representative applications. Section 3 presents the theoretical framework and methodology behind the QAE model used and description of the dataset. Section 4 describes the experimental setup and evaluation methodology for medical imaging and anomaly detection tasks. Section 5 discusses the experimental results, and Section 6 concludes the paper. Additional analyses and results are provided in the Appendix Section 7.

## 2 Autoencoders: Background

This section reviews classical and quantum autoencoders (QAEs), outlining their theoretical foundations and recent research.

### 2.1 Quantum Autoencoder (QAE)

A generalized quantum autoencoder is shown in Fig. 1(b) [48]. In the case of a CAE (shown in Fig. 1(a)) [132-135], the function is to optimize its parameters across an input $x$ drawn from a training dataset to learn to reconstruct the original input. The encoder maps the input from an $(n + k)$-dimensional space into a compressed latent representation of dimension $n$, effectively discarding $k$ dimensions of information during the encoding process. When the input data are represented as binary string, this $(n + k)$-dimensional input equivalently viewed as an $(n + k)$-bit string $x$ training set of data, and try to reconstruct $x$; in order to reconstruct $x$, the CAE circuit has to discard some $k$-bits of data. Therefore, a CAE, once successfully trained, can approximately reconstruct $x$ as an output, has $n$-bits as encoded compression of string $x$ and hence, "gets trained" to encode/decode information in close proximity of the original dataset.

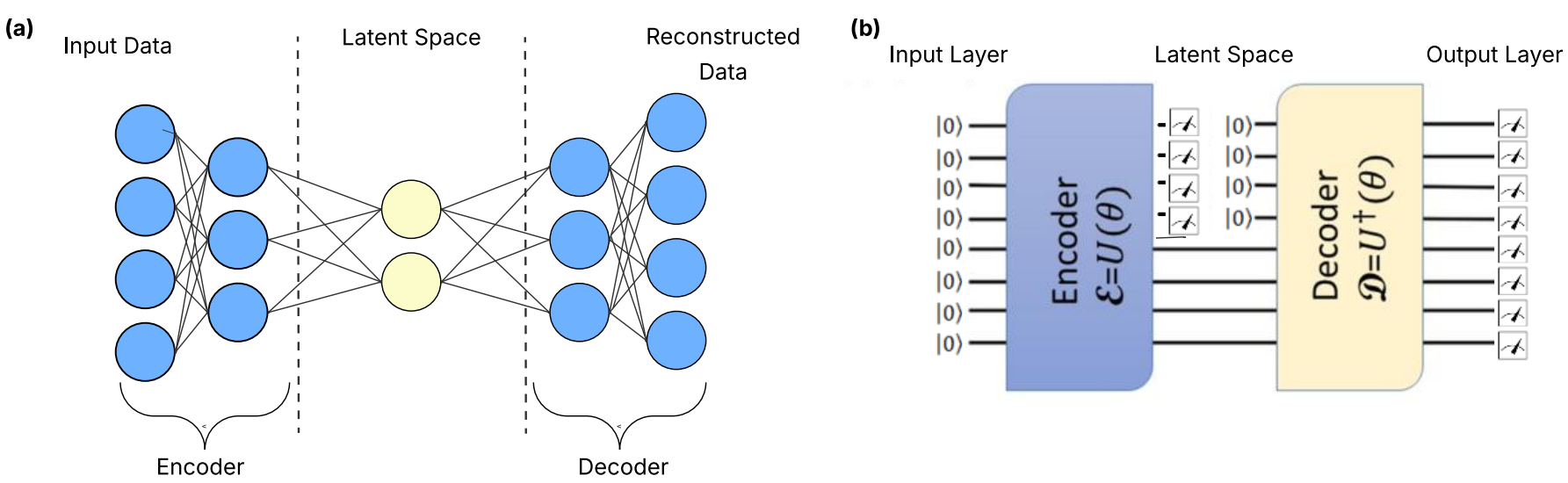

**Fig. 1.** General structure of (a) Classical and (b) Quantum autoencoders

The concept of QAE [48] was inspired by that of the CAE [36-46]. Typically, *quantum circuit learning* (QCL) is used to build a QAE. Like its classical counterpart CAE, the goal of QAE is compression and reconstruction of input data. QAE is also constructed of an encoder $\mathcal{E}$, a latent space, and a decoder $\mathcal{D}$. Whereas in a CAE parts of the feature space can be discarded during dimensionality reduction, its quantum counterpart must preserve the entire Hilbert space because quantum operations are implemented by unitary transformations. To achieve the desired dimensionality reduction, the encoder transfers redundant information to ancillary ('trash') qubits. These qubits are then traced out, thereby reducing the effective dimensionality of the representation.

In the case of QAE, the encoder and decoder are implemented as sets of unitary transformations represented by gates in parametric quantum circuits (PQC). As depicted in Fig. 1(b), $\theta$ represents the trainable parameters of the unitary transformation $U(\theta)$. The input classical data is encoded into a quantum state $|\psi\rangle$ comprising of $n$ qubits, which is then compressed into the quantum latent space by optimizing the ansatz $U(\theta)$, and eventually reconstructed to a close approximation of the original input after passing it through the decoder $\mathcal{D}$. Compression is achieved by application of a PQC on $|\psi\rangle$. Reconstruction is achieved by inverting the encoder unitary to $U^{\dagger}(\theta)$. The set of qubits used to compress the data are called "latent qubits", and the discarded qubits that are traced out are called "trash qubits". The *learned representation* is given by the remaining subsystem after the trash qubits are traced out. New initialized qubits are prepared and used to implement the final decoding task via $\mathcal{D}$ and then the input and output states are compared.

The learning objective of the quantum encoder is to learn parameterized unitary transformations that preserve the input information as it is processed through the lower dimensional latent space.

During training, a quantum decoder is attached to the latent space to regularize the learned representation and encourage information-preserving compression, The decoder mirrors the encoder structure and attempts to reconstruct the original quantum state. Although the decoder is not used during inference, it enables an auxiliary auto-encoding objective that stabilizes training and discourages trivial representations. Our QAE formulation is described in detail in Section 3.2.

# 3 Methodology

This section discusses the methods, and the experimental settings of our approach and is divided as follows: section 3.1 discusses the Brain MRI DICOM data, and its characteristics; section 3.2 describes the general Romero *et al*. [48] type QAE applied to the Brain MRI data, the quantum circuit design, training and reconstruction evaluation, compression-reconstruction trade-off methodology, parameter description, evaluation metrics for classical baseline, anomaly detection metrics, and experimental details.

## 3.1 Brain MRI Data: Application of QAE

Unsupervised anomaly detection plays a critical role in medical imaging, where pathological patterns are often rare, heterogeneous, and lack reliable labels. Quantum autoencoders [48] offer a method complementary to classical frameworks in which compression can be explicitly enforced at the level of quantum states through auxiliary qubits, allowing anomalous inputs to be identified by their inability to be efficiently compressed. In this section, we apply a Romero *et al.* type [48] quantum autoencoder to medical imaging–derived greyscale brain MRI DICOM data [113] to investigate compression-driven anomaly detection and to analyze how its internal circuit structure supports this task. The novelty of this work is the formulation of *brain MRI anomaly detection as a compression-driven task using a quantum autoencoder*, where anomaly scores are derived from *trash-qubit compression behavior* rather than reconstruction error alone. In addition, the study provides a *mechanistic analysis of encoder–decoder asymmetry* and demonstrates *spatially localized anomaly heatmaps*, making the proposed QAE both interpretable and practically relevant for medical-image anomaly screening.

By examining reconstruction fidelity, parameter structure, and layer-wise energy allocation, we aim to provide insight into how quantum autoencoders allocate representational capacity in practical anomaly detection settings. We next detail the quantum autoencoder architecture, training objective, and analysis framework used to study compression-driven anomaly detection in the brain MRI DICOM [113] dataset.

**Brain MRI DICOM data description.** We ran an anomaly detection exercise utilizing QAE with high-resolution greyscale Brain MRI data from the Digital Imaging and Communications in Medicine (DICOM), which serves as a technical standard for the digital storage and transmission of medical images. The dataset, which is publicly available, consists of .dcm files containing *MRI scans of the brain* of persons with a normal brain and has been made available for studies on anomaly classification and detection. The images are *labeled* by the doctors and accompanied by a report in PDF-format.

The dataset includes seven studies as shown in **Table 1** and are made from the different angles which provide a comprehensive understanding of a normal brain structure and useful in training brain anomaly classification algorithms. An example of the input DICOM data for a tumorous brain is shown in **Fig. 2**, and the anomaly detection is shown in **Fig. 7** and **Fig. 8** of main the main paper and **Fig. 14** of the Appendix**.**

**Table 1.** DICOM Brain Data [113]

| Directory | Normal | Tumor/Anomaly |
|---|---|---|
| SE000001 | 9 | 27 |
| SE000002 | 50 | 27 |
| SE000003 | 27 | 34 |
| SE000004 | 27 | 28 |
| SE000005 | 36 | 50 |
| SE000006 | 27 | 09 |
| SE000007 | 24 | 24 |

The publicly available dataset was downloaded from Hugging Face [113]. The dataset is split into seven directories of unequal numbers of files in categories of “normal” and “tumor”. The distinct and greyscale dataset were compiled from the different directories/classes and analysed as a single class. The “tumor” dataset is used as “anomaly” to the “normal”. Table 1 shows the data split.

Brain MRI datasets on Hugging Face [113] vary widely in size and image resolution, depending on the specific dataset and whether it offers a full or sample version. The overall size of brain MRI datasets is highly variable, with some professional datasets containing millions of images, while samples available on Hugging Face often provide smaller subsets of data for evaluation purposes. Downloadable sample files are typically in the range of $10-40$ MB, the files are in *.dcm* (DICOM) format, as well as in .jpg format. Individual DICOM file sizes also vary depending on the sequence and compression, with a single uncompressed scan potentially exceeding 30 MB, while a single T1w MRI sample might be around 50 MB.

The *resolution of images* within these datasets were also found to be *diverse*, as it is determined by the specific MRI scanner and acquisition parameters. Common resolutions and parameters include:

- **Pixel Resolution (Matrix Size):** Typical matrix dimensions for MRI data range from $128 \times 128$ to $512 \times 512$ pixels or even higher. By inspection, our dataset had resolution spanning between $128 \times 128$ to $256 \times 256$ to $256 \times 333$. Resolution was found to be mixed and ununiform.
- **Bit Depth:** DICOM files use higher bit precision than standard images, in this case, 12-bit per pixel precision.

The input data with mixed resolution was resized to $128 \times 128$ for a QCNN patch size of $2 \times 2$, mapped into 4-qubits. The training was run for a batch size of 32 for 10-epochs, using Adam optimizer.

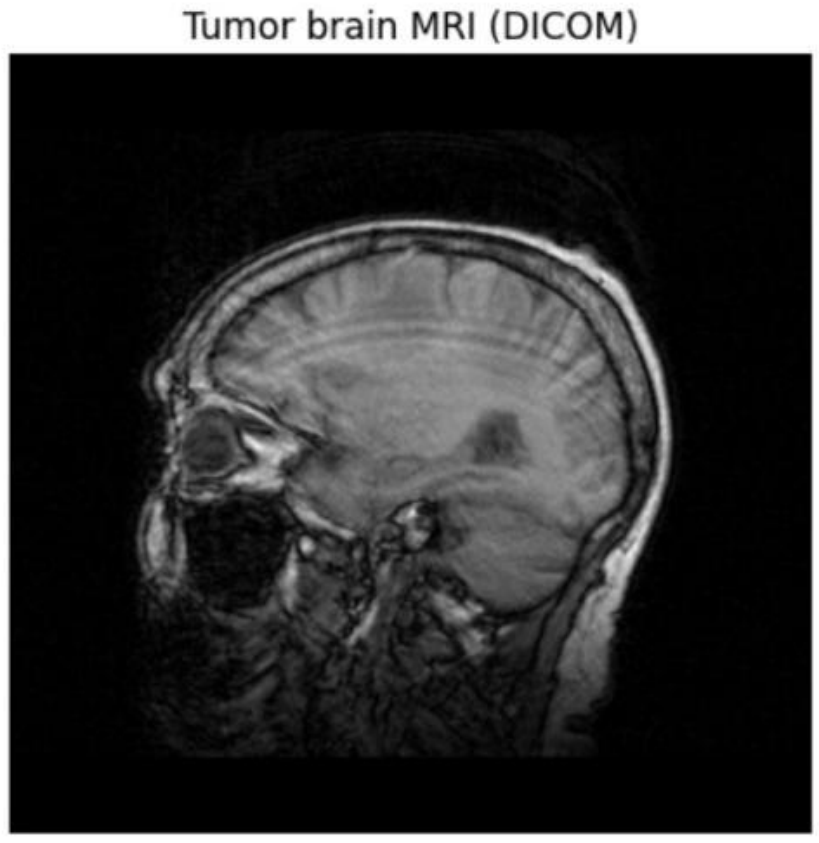

**Fig. 2.** Sample DICOM image for a tumorous brain

We next describe the quantum autoencoder architecture, training objective, and evaluation methodology used to analyze compression-driven anomaly detection in medical imaging–derived data.

### 3.2 Quantum Autoencoder Architecture for the Brain MRI Data

We employ a four-qubit quantum autoencoder (QAE) adapting from Romero *et al.* [48, 118-121] comprising two logical qubits and two auxiliary (trash) qubits, following the general framework of quantum state compression via auxiliary subsystems [127]. We employ angle encoding in this study due to the simpler image depth, shallow circuit depth, and required 4-qubit architecture for the greyscale images. Classical input vectors $\boldsymbol{x} \in \mathbb{R}^{\mathbf{4}}$, normalized to $[0, \pi]$, are embedded into quantum states using *angle encoding*. The QAE circuit consists of a parameterized encoder followed by a decoder with identical topology but independent parameters (**Fig. 3**). All experiments are performed using a state-vector simulator. By examining reconstruction fidelity, parameter structure, and layer-wise energy allocation, we aim to provide insight into *how* quantum autoencoders allocate *representational capacity in practical anomaly detection settings*.

In this work:

- We provide a **principled characterization of the compression–reconstruction trade-off** in a quantum autoencoder trained on normal medical imaging data (cf. **Fig. 4**).
- We present an **interpretable, parameter-level analysis** of the learned encoder and decoder circuits, revealing structured asymmetries in parameter usage and layer-wise energy allocation that explain the observed trade-off (cf. **Fig. 6**- **Fig. 8**).
- We analyze the layer-wise parameter energy with histograms to quantify how parameters are distributed across quantum circuit layers (cf. **Fig. 6**, **Fig. 7** and **Fig. 9**).
- We showcase anomaly heatmap generated from patch-level QAE anomaly scores (cf. **Fig. 8**).
- We contextualize these findings through **carefully matched classical baselines**, including principal component analysis and a classical autoencoder with equivalent latent dimensionality.

**QAE Formulation.** A quantum autoencoder (QAE) is a parametrized quantum circuit that learns to compress input states into a lower-dimensional latent subsystem while discarding redundant information into a designated **trash subsystem**. The original QAE formulation of Romero *et al.* [48] introduced this idea for compressing families of quantum states by training a unitary so that the trash qubits are mapped to a fixed reference state, typically $|0\ldots0\rangle$. In hybrid quantum machine learning, the same idea can be adapted to classical data by first embedding a classical feature vector into a quantum state and then optimizing the circuit parameters with a classical optimizer [15]. The *embedding step* can be interpreted as a nonlinear feature map into Hilbert space, and differentiable frameworks such as PennyLane [57] enable gradient-based optimization of such hybrid models.

Let $\mathbf{x} \in \mathbb{R}^d$ denote an input feature vector. In this work, each $2 \times 2$ image patch is flattened to $d = 4$ features and rescaled to $[0, \pi]^4$. The classical-to-quantum embedding is implemented by angle encoding,

$$|\psi_{\text{in}}(\mathbf{x})\rangle = U_{\text{feat}}(\mathbf{x})|0\rangle^{\otimes n}, \qquad U_{\text{feat}}(\mathbf{x}) = \prod_{i=1}^{d} R_y(x_i),$$

where $n = 4$ is the number of qubits and $R_y(x_i)$denotes a single-qubit rotation. This type of embedding is consistent with the general feature-map view of quantum machine learning, in which classical inputs are mapped into quantum states in a Hilbert-space representation. The QAE then applies a variational encoder $U_\theta$followed by a variational decoder $V_\phi$,

$$|\psi_{\text{out}}(\mathbf{x};\theta,\phi)\rangle = V_\phi\, U_\theta\, U_{\text{feat}}(\mathbf{x})|0\rangle^{\otimes n}.$$

In our implementation, both $U_\theta$ and $V_\phi$ are composed of repeated layers of single-qubit $R_y$ rotations and ring-CNOT entangling operations. The qubit register is partitioned into a latent subsystem $L$ and a trash subsystem $T$. Compression is successful when the reduced state on $T$ is close to the reference state $|0\cdots0\rangle_T$. This is the key operational principle of the QAE. Our QAE circuit is shown in **Fig. 3** below.

Let $p(\mathbf{z}|\mathbf{x};\theta,\phi)$ denote the probability of observing computational-basis bitstring $\mathbf{z}$after measurement. The probability that all trash qubits are reset to zero is

$$p_{\text{good}}(\mathbf{x}) = \sum_{\mathbf{z}:\,\mathbf{z}_T=\mathbf{0}} p(\mathbf{z}|\mathbf{x};\theta,\phi).$$

This quantity is the measurement-based analogue of the trash-state fidelity used in the original QAE formulation. For a *single trash qubit* $t$, the same quantity can be computed more efficiently from a Pauli-$Z$ expectation value,

$$p_{\text{good}}(\mathbf{x}) = \frac{1+\langle Z_t\rangle}{2}.$$

In practice, this relation enables a faster encoder-only inference mode for anomaly scoring. Training is performed on normal patches only. For samples $\mathbf{x} \sim \mathcal{D}_N$, the basic compression loss is

$$\mathcal{L}_N(\theta,\phi) = \mathbb{E}_{\mathbf{x}\sim\mathcal{D}_N}[1 - p_{\text{good}}(\mathbf{x})].$$

Minimizing $\mathcal{L}_N$ encourages the model to map normal inputs into states that are easily compressible, i.e. states whose trash qubits are close to the fixed reference. Similar trash-qubit compression metrics have also been used in later QAE-based anomaly-detection settings.

To avoid a trivial solution in which broad classes of inputs achieve high trash fidelity, we introduce **pseudo-anomalies** $\tilde{\mathbf{x}} = \mathcal{C}(\mathbf{x})$ through feature corruption or dropout and penalize high compressibility on these samples. An equivalent contrastive objective is

$$\mathcal{L}(\theta,\phi) = \mathbb{E}_{\mathbf{x}\sim\mathcal{D}_N}[1 - p_{\text{good}}(\mathbf{x})] + \alpha\,\mathbb{E}_{\tilde{\mathbf{x}}\sim\mathcal{C}(\mathcal{D}_N)}[p_{\text{good}}(\tilde{\mathbf{x}})],$$

where $\alpha > 0$ controls the strength of the pseudo-anomaly term. Thus, the model is encouraged to assign **high compression fidelity** to normal inputs and **lower compression fidelity** to corrupted inputs. For anomaly detection [48, 102], the natural compression-based score is

$$a(\mathbf{x}) = 1 - p_{\text{good}}(\mathbf{x}),$$

so that poorly compressed inputs receive larger anomaly scores. In our experiments, however, validation AUC indicated that the empirically best monotone score direction was $p_{\text{good}}$ itself; accordingly, a calibrated score $s(\mathbf{x})$was obtained by applying a power transform

$$s_\tau(\mathbf{x}) = s(\mathbf{x})^\tau, 0 < \tau \leq 1,$$

where $\tau$is tuned on a validation subset. This calibration is monotone and therefore preserves within-patch ranking, while improving the dynamic range for subsequent slice-level aggregation.

Finally, because tumours *typically occupy only a limited fraction of a brain MRI slice*, patch scores are aggregated using a **top-**$k$ rule rather than uniform averaging. For slice $j$ with patch scores $\{s_{j,i}\}_{i=1}^{N_j}$,

$$S_j^{(k)} = \frac{1}{K_j} \sum_{i \in \text{TopK}_j} s_\tau\,(\mathbf{x}_{j,i}), K_j = \max\,(1, \lceil kN_j \rceil),$$

where $\text{TopK}_j$ indexes the highest-scoring $K_j$ patches. This yields a slice-level anomaly score $S_j^{(k)}$, which is used for ROC analysis and threshold-based decision support (please refer to Section 3.6 for more details).

### 3.3 QAE Circuit Design

Each input component is encoded using single-qubit rotations, producing an input-dependent target quantum state $|\psi_t(x)\rangle$. The encoder and decoder each consist of two parameterized layers, where each layer applies single-qubit $RY$and $RZ$ rotations to all qubits followed by a nearest-neighbor CNOT entangling block. The circuit structure is representative of variational quantum algorithms used in quantum machine learning [22, 48, 72, 127]. The quantum circuit for this experiment was generated during code execution, and is shown in **Fig. 3**.

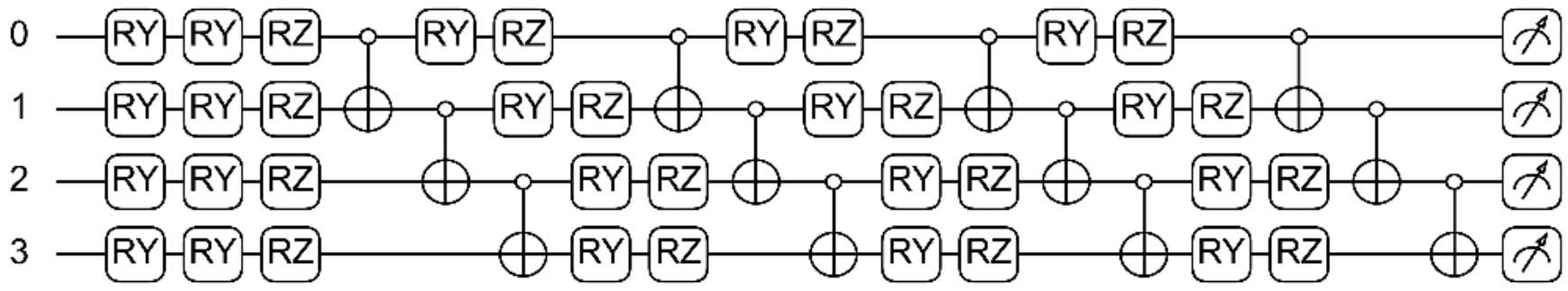

**Fig. 3.** QAE circuit used in this study. Produced with PennyLane [57] as output of the code run.

### 3.4 Training and reconstruction evaluation

The quantum autoencoder is trained on normal image patches using a trash-qubit objective that encourages the auxiliary qubits to occupy $|00\rangle$ computed from the measured output probabilities. Reconstruction quality is evaluated independently using a fidelity-based metric: for each input sample $x_i$, the target state $|\psi_{\text{t}}(x)\rangle$ is defined as the state produced by the encoding circuit alone, the target state $|\psi_{\text{t}}(x)\rangle$ is defined as the state produced by the angle-encoding circuit alone, and the reconstructed state $|\psi_{\text{o}}(x;\theta)\rangle$ after the full encoder–decoder circuit. Pure-state fidelity is $F(x) = |\langle\psi_{\text{t}}(x) \mid \psi_{\text{o}}(x;\theta)\rangle|^2$, and the reconstructed state $|\psi_{\text{out}}(x)\rangle$ is obtained after applying the full encoder–decoder circuit.

*Reconstruction fidelity* is computed as $F(x) = |\langle\psi_{\text{t}}(x) \mid \psi_{\text{out}}(x)\rangle|^2$, and *reconstruction loss* is defined as *infidelity* $1 - \mathbb{E}[F]$, where $\mathbb{E}$ is the expectation value describing average reconstruction fidelity over the data distribution. In our QAE set-up, the average over a dataset (or a batch) is given by,

$$\mathbb{E}[F] \approx \frac{1}{N}\sum_{i=1}^{N} F(x_i),$$

which gives infidelity as

$$\text{Infidelity} = 1 - \mathbb{E}[F] = 1 - \frac{1}{N}\sum_{i=1}^{N} F(x_i)$$

To reduce computation, target states are precomputed once and reused during training without approximation. Uncertainty bands in trade-off plots are reported as the *standard error of the mean* (SEM) across mini-batches within each epoch. Fidelity varies per sample, as each input patch produces a different quantum state. Hence, a *single scalar* is required to track training progress, trade-offs during plotting and comparison of epochs. The training loss which is given by the trash loss, is also an expectation:

$$L_{trash} = \mathbb{E}[1 - \mathrm{P}(trash = |00\rangle)]$$

Hence, using $\mathbb{E}[F]$ allows to preserve both training loss and reconstruction loss and give an output on the same statistical footing. Expectation values are specifically taken over the empirical data distribution and approximated by finite sample average. In our implementation, this has been done over mini-batch during training and over validation set for reporting.

**Compression–Reconstruction Trade-off and Knee Selection:** Training induces a trade-off between compression (trash loss) and reconstruction (infidelity), analogous to rate–distortion behavior in information theory [128]. This trade-off is analyzed using Pareto curves plotting trash loss against infidelity across training epochs (**Fig. 13**). A knee operating point is identified using a *curvature-based L-curve criterion*, defined as the epoch with maximum perpendicular distance from the line connecting the first and last points in normalized loss space. This knee corresponds to a balanced regime beyond which further compression yields diminishing reconstruction returns.

### 3.5 Experimental Parameter

The experimental parameters are listed in 2 below.

**Table 2.** Parameter values and Interpretation: Brain MRI data treatment by the QAE

| Hyperparameter | Value | Meaning |
|---|---|---|
| *n_qubits* | 4 | Size of quantum register |
| *latent_qubits* | 2 | Target quantum compression dimension |
| *patch_size* | (2, 2) | Converts image patches → 4 qubits |
| *learning_rate* | 2.00E-02 | Adam optimizer step size |
| *epochs* | 30 | Number of epochs |
| *batch_size* | 32 | Patches per quantum batch |

Table 2 lists the parameters specifically for the QAE model. Our QAE model carries no classical layers – both the encoder and decoder utilises quantum circuit learning [87, 95].

**Parameter Structure and Layer Interpretation:** Learned parameters are analyzed using **encoder and decoder heatmaps and distributions** (**Fig. 15**). Parameters are indexed by layers $L0$–$L3$ and qubits $q0$–$q3$, where:

- **L0**: first-layer $RY$ rotations,
- **L1**: first-layer $RZ$ rotations,
- **L2**: second-layer $RY$ rotations,
- **L3**: second-layer $RZ$ rotations.

This representation enables direct interpretation of how early and late circuit layers contribute differently to compression and reconstruction.

**Parameter Structure and Layer-wise Energy:** Learned parameters are analyzed using encoder and decoder heatmaps (Fig. 14), indexed by layers $L0$–$L3$, corresponding to $RY$ and $RZ$ rotations in the first and second circuit layers. Layer-wise parameter energy is quantified using the L2 norm of rotation angles,

$$E_{\ell,q} = \sqrt{\theta_{\ell,q}^2 + \phi_{\ell,q}^2},$$

averaged across qubits and reported with SEM. Normalized energy allocation reveals pronounced encoder–decoder asymmetry, with compression localized in later encoder layers (Fig. 3).

### 3.6 Experimental setup

Experiments were performed on publicly available brain MRI DICOM datasets containing normal and tumorous slices. After per-slice normalization and resizing, each image was decomposed into overlapping $2 \times 2$ patches, which were flattened into 4-dimensional vectors and angle-encoded into a four-qubit quantum autoencoder. The variational circuit consisted of repeated $RY$ rotation layers and ring-CNOT entanglement in both the encoder and decoder. Training was carried out in a normal-only setting using a hybrid PennyLane–PyTorch implementation on a noiseless state-vector simulator. A contrastive corruption term based on feature dropout was used to discourage trivial compression. Full experimental set-up including data evaluation methods and software versions are shared in Section 7.1 of the Appendix of this paper.

### 3.7 Evaluation Metrics

The AUC metric from the multi-class dataset is calculated using softmax activation function. The expectation values were converted into probabilities of classes to ensure the output sum resulted to 1 to represent probabilities for each class. To calculate the ACC, the index of the maximum value was extracted from each row of the expectation values vector and compared with the original labels to give,

$$\text{ACC} = \frac{\text{Correct predictions (number)}}{\text{Total samples (number)}}$$

Anomaly scores are defined as $1 - \Pr(trash = |00\rangle)$, where higher scores indicate greater deviation from the learned normal manifold. Performance is evaluated using **ROC–AUC** as the primary, threshold-independent metric [123-126]. For operational assessment, **accuracy** and **F1 score** are reported at the Pareto knee operating point identified from the compression–reconstruction trade-off [128], with thresholds selected via Youden's J [129] statistic on validation data.

For evaluation, confusion matrices, AUC/ROC, loss, accuracy curves with errors, F1 scores, Accuracy (ACC), precision and recall for each dataset were produced. We define the metrics as shown in **Table 3** below.

**Table 3.** Definition of Metrics

| Metrics | Equation |
|---|---|
| Accuracy | $\frac{\text{TP} + \text{TN}}{\text{TP} + \text{TN} + \text{FP} + \text{FN}}$ |
| Precision | $\frac{\text{TP}}{\text{TP} + \text{FP}}$ |
| Recall | $\frac{\text{TP}}{\text{TP} + \text{FN}}$ |
| F1 Score | $\frac{2 \times \text{Precision} \times \text{Rec}}{\text{Precision} + \text{Recal}}$ |

**Evaluation of patch-level discrimination and slice-level anomaly detection.** Patch-level scores were used to assess local anomaly separability, but because patches from the same MRI slice are not statistically independent, inferential uncertainty was estimated by resampling at the slice level (which is a common practice for medical MRI images) [51, 124] rather than at the patch level. Hence, patch-level anomaly scores were first computed for all validation patches. Because multiple patches originate from the same MRI slice, patch-level observations were not treated as independent samples for inferential testing. Instead, patch-level discrimination was used primarily as a descriptive analysis of local separability, whereas formal uncertainty estimates were obtained by **resampling at the slice level**.

To characterize patch-level separation between normal and anomalous samples, we computed the receiver operating characteristic area under the curve (ROC–AUC) together with nonparametric distributional distances. In particular, the **Kolmogorov–Smirnov (KS) statistic** [136] was used to quantify the maximum discrepancy between empirical score distributions, and the first **Wasserstein distance** [137] was used to measure the transport distance between the two distributions. For *qubit-wise latent analyses*, the same statistics were computed on the corresponding per-qubit $L_2$ magnitude distributions. These quantities were interpreted as measures of distributional separation rather than as independent evidence detached from slice-level evaluation.

To obtain slice-level anomaly scores, patch scores within each slice were aggregated using **top-$k$ pooling**. For a slice $j$ with patch scores $\{s_{j,i}\}_{i=1}^{N_j}$, the slice score was defined as

$$S_j^{(k)} = \frac{1}{K_j} \sum_{i \in \mathrm{TopK}_j} s_{j,i}\,, \qquad K_j = \max\left(1, \lceil kN_j \rceil\right),$$

where $\mathrm{TopK}_j$ denotes the indices of the largest $K_j$ patch scores. This aggregation was motivated by the expectation that tumorous regions occupy only a limited subset of image patches, so the anomaly signal is spatially sparse rather than uniformly distributed across the slice.

Slice-level discrimination was evaluated using ROC curves and slice-level AUC. Statistical significance above chance level was assessed using a *nonparametric permutation test*: slice labels were randomly permuted $B$ times (e.g., $B = 10{,}000$), the AUC was recomputed for each permutation, and the empirical $p$-value was defined as

$$p = \frac{1 + \sum_{b=1}^{B} \mathbf{1}\left(\mathrm{AUC}^{(b)} \geq \mathrm{AUC}_{\mathrm{observed}}\right)}{B + 1}.$$

Uncertainty in slice-level AUC was estimated using *stratified bootstrap resampling over slices*, preserving the class balance in each bootstrap replicate. When comparing top-$k$ aggregation with alternative pooling strategies such as mean or max pooling, significance of the improvement was assessed using the *paired bootstrap difference in*

*AUC*, computed on the same bootstrap resamples. An improvement was considered supported when the corresponding 95% confidence interval for ΔAUC excluded zero.

Decision thresholds for reporting slice-level accuracy were selected on a separate validation or tuning subset using either the *Youden index* [129] or a fixed specificity target, and then held fixed for final evaluation.

### 3.8 Model pipeline

We share the *full model pipeline* used in this study in **Fig. 15** in Appendix.

## 4 Results and Discussions

In this section we showcase anomaly detection utilizing angle encoded QAE leveraging publicly available mixed high-resolution, greyscale Brain MRI sample dataset from Hugging Face [113].

As described in Section 3, we ran an anomaly detection exercise utilizing angle encoding driven QAE architecture to analyse high-resolution greyscale Brain MRI sample data from the *Digital Imaging and Communications in Medicine* (DICOM) [113], which serves as a technical standard for the digital storage and transmission of medical images. The data from seven directories were collapsed into two datasets: 200 samples of distinct "normal" class and 195 samples of "anomaly" dataset with diagnosed tumors. The publicly available data was downloaded from *Hugging Face* [113], preprocessed to $128 \times 128$, normalized, and run for 10 epochs at a batch size of 32. Anomaly scores and plots, accuracy and AUC/ROC curves, loss, accuracy and validation curves, parameters, normal-anomaly score distributions, and visual anomaly heatmap to identify tumorous growths were generated.

Total *trainable parameters* in quantum part of the workflow were 32. Given the resolution of the DICOM brain MRI dataset was highly diverse and unbalanced, varying between $128 \times 128$ and $256 \times 333$, the accuracy and AUC obtained by this first ever known run though a QML pipeline was promising. Our reported QAE based classification performance of a final AUC with an average peak AUC of 81.3 % and average accuracy of 75.22% is an improvement over directly compared classical results (**Fig. 11**- **Fig. 13**). The QAE results vary in AUC scores between $\sim$41.2% $\pm$ 0.41 for SE000004, and SE000006 and between 76.4% to 82.5% for SE000002. We report a maximum slice-level ROC of 95% (**Fig. 10** in Appendix) The F1 score at the knee was found to be 0.70.

### 4.1 Comparison with Classical Baselines

To contextualize performance, the quantum autoencoder is compared against two classical baselines: principal component analysis (PCA) with four components and a classical autoencoder with matched latent dimension. All models are trained on normal data only and evaluated using identical anomaly scores and metrics.

Using the knee operating point identified from the compression–reconstruction trade-off, anomaly detection performance is evaluated using trash-based scores. The quantum autoencoder achieves stable separation between normal and anomalous samples, reflected in consistently higher ROC–AUC values than classical baselines. In contrast, both classical autoencoder and PCA baselines exhibit heavily overlapping score distributions and thresholds that collapse toward zero, limiting their practical utility. **Fig. 11**-**Fig. 13** in the Appendix shows the comparison plots for the ROC curves and AUC obtained between QAE, Classical and PCA methods and associated analyses. **Table 4** below shows a comparison of the values.

**Table 4.** Comparison of anomaly detection performance and mechanisms

| Model | Slice-ROC | AUC (≈) | Score distribution | Threshold behavior | Mechanism |
|---|---|---|---|---|---|
| QAE | 0.95 | 0.781–0.813 | Normal tightly near zero; anomalies exhibit a heavy right tail | Non-zero, stable threshold | Compression via trash qubits |
| Classical AE | 0.41 | ~0.63–0.64 | Strong overlap; narrow dynamic range | Threshold ≈ 0 | Reconstruction error |
| PCA | - | ~0.56–0.58 | Almost fully overlapping | Threshold ≈ 0 | Linear variance truncation |

### 4.2 Compression-Reconstruction Trade-Off: Pareto.

**Fig. 4** shows the Pareto plot for the QAE with an explicit compression–reconstruction trade-off, analogous to rate–distortion theory. This allows principled operating-point selection, which classical methods do not provide.

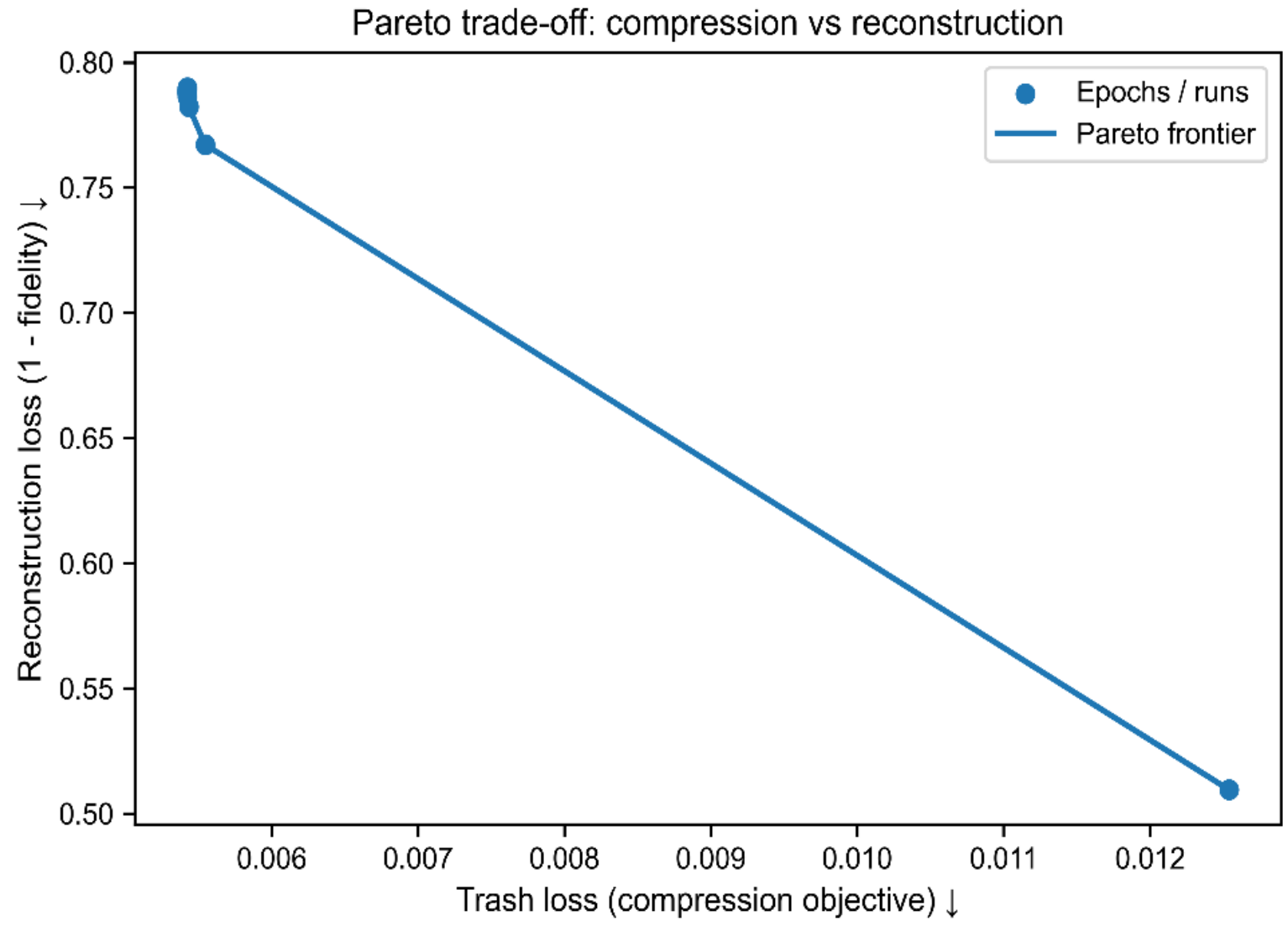

**Fig. 4.** Trash loss (compression objective) is plotted against reconstruction loss (infidelity) across training epochs. Points correspond to individual epochs. The knee of the curve identifies a balanced operating regime used for downstream anomaly detection evaluation

**Fig. 4** shows that for the QAE normal scores are sharply concentrated near zero; Anomalies exhibit a long right tail; The selected threshold is *small but non-zero* ($\approx$ 0.002–0.003). The Pareto frontier highlights the monotonic trade-off between increased compression and degraded reconstruction fidelity. A comparison of anomaly detection performance and underlying mechanisms across models is provided in **Table 4**.

The plots indicate that the QAE learns to **compress normal data efficiently** into the latent subspace, driving trash qubits close to $|00\rangle$. Anomalies resist this compression, producing higher trash loss. This yields a clear score asymmetry, which is *typical* for unsupervised anomaly detection and verifies our QAE application performance. These results demonstrate that compression-driven anomaly detection provides a more informative detection signal than reconstruction error alone.

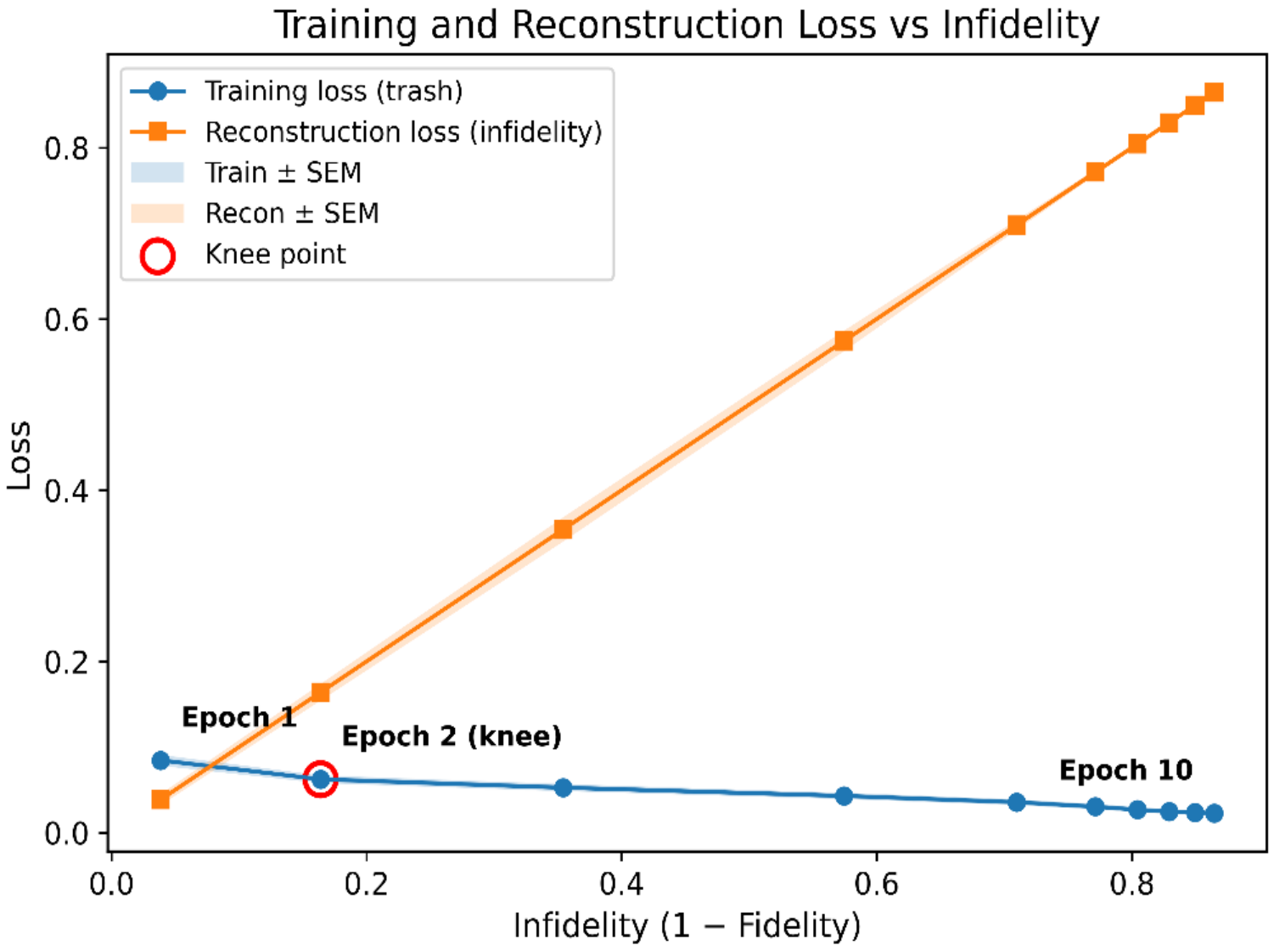

**Fig. 5.** The training (trash) loss & reconstruction loss against infidelity for the QAE.

## 4.3 Compression–Reconstruction vs. Infidelity

**Fig. 5** illustrates the relationship between compression and reconstruction during training of the quantum autoencoder. As training progresses, the trash loss decreases monotonically (blue curve), indicating increasingly effective compression of information into the latent subspace, while the reconstruction loss (infidelity) increases (orange curve), reflecting degradation of reconstruction quality as more information is discarded.

The curve in **Fig. 5** reveals a smooth compression–reconstruction trade-off characteristic of autoencoding architectures, rather than unstable or oscillatory training dynamics. The shaded regions denote ±SEM across mini batches, demonstrating that both losses evolve consistently across training. A knee point is identified at Epoch 2 (circled), corresponding to a regime where a substantial reduction in trash loss is achieved with only a modest increase in infidelity. Beyond this point, further compression yields diminishing returns, as relatively small reductions in trash loss lead to disproportionately large increases in reconstruction loss. This knee therefore defines a principled operating point for downstream anomaly detection evaluation, and all threshold-dependent metrics reported subsequently are computed at this operating regime.

## 4.4 Layer-wise Parameter Energy

**Fig. 6** quantifies the qualitative parameter asymmetry observed in **Fig. 4** and **Fig. 5** using layer-wise L2 parameter energy.

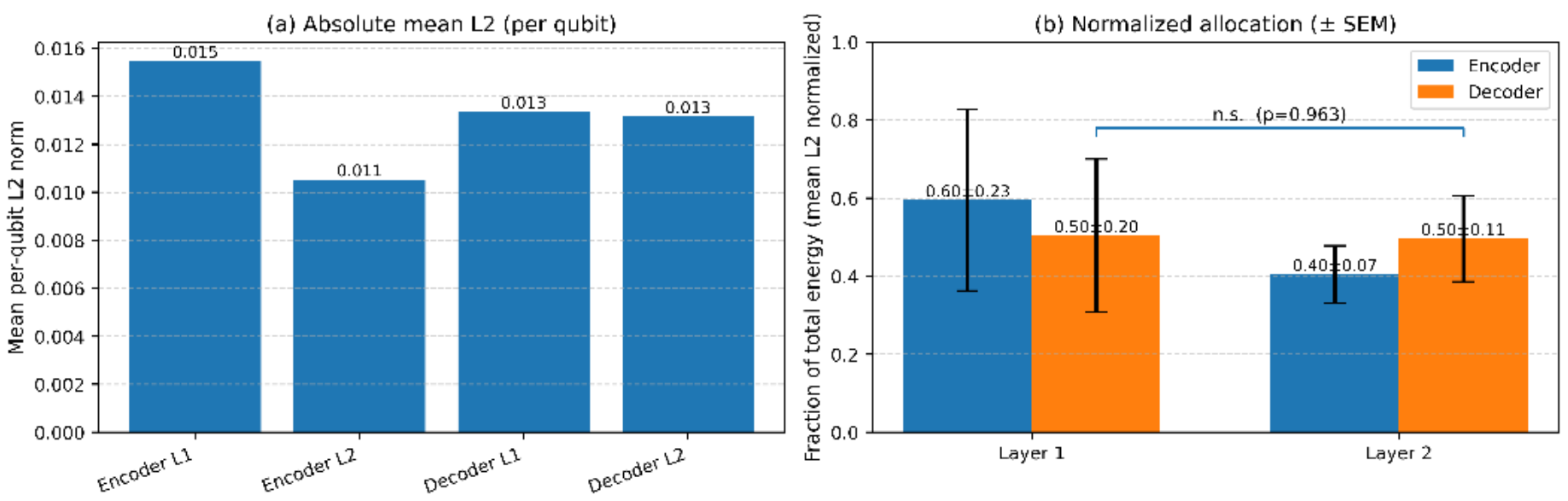

**Fig. 6.** Layer-wise **parameter energy allocation.** (a) Absolute mean per-qubit L2 norms of encoder and decoder parameters. (b) Normalized fraction of total parameter energy allocated to each layer, reported as mean ± SEM across qubits. While absolute magnitudes are comparable, the encoder concentrates a larger fraction of its energy in the first parameterized layer, whereas the decoder distributes energy more uniformly. No significant difference is observed in total normalized energy allocation between encoder and decoder layers (n.s., $p = 0.963$)

**Fig. 6(a)** shows the absolute mean per-qubit L2 norms for encoder and decoder layers, indicating comparable overall parameter magnitudes across the circuit. To disentangle magnitude from allocation, **Fig. 6(b)** reports the normalized fraction of total parameter energy per layer, together with SEM across qubits. The encoder exhibits a clear bias toward the first parameterized layer, whereas the decoder distributes its energy more uniformly across layers. Importantly, the total normalized energy does not differ significantly between encoder and decoder layers (n.s., $p = 0.963$), indicating that compression arises from structured redistribution of parameters rather than increased expressivity. This layer-specific energy allocation provides a quantitative explanation for the encoder–decoder asymmetry and supports the interpretation that compression is localized within the encoder, consistent with the trade-off behavior observed in **Fig. 5**.

### 4.5 Encoder–Decoder Parameter Structure

To understand the origin of the observed trade-off, we examine the learned parameter structure of the encoder and decoder circuits (**Fig. 7**). Encoder parameters exhibit pronounced layer- and qubit-dependent patterns, with structured activations and a broad distribution of values. In contrast, decoder parameters are sharply concentrated near zero, with only localized deviations.

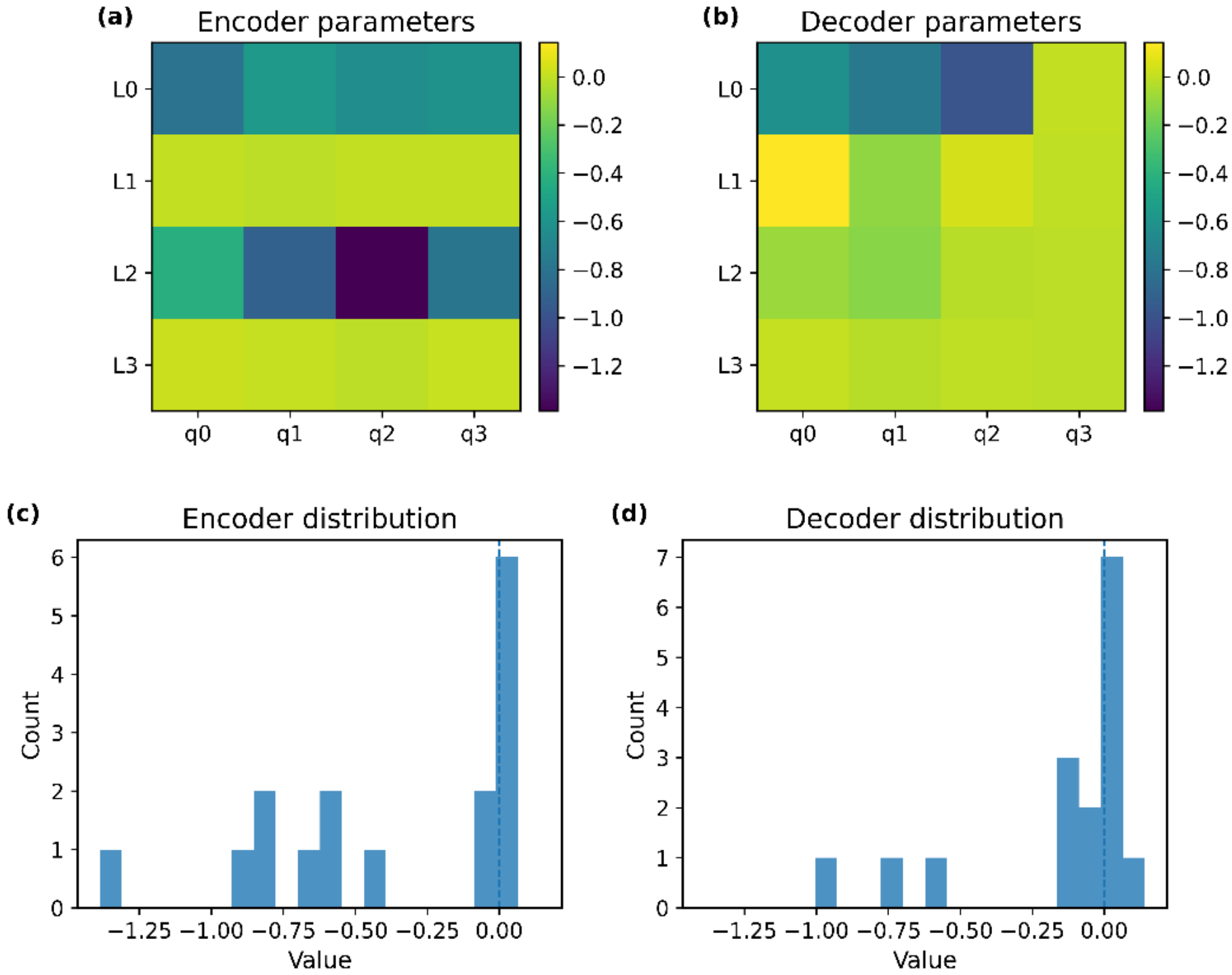

**Fig. 7.** Encoder–decoder **parameter structure.** (a,b) **Heatmaps** of learned encoder and decoder parameters across layers (**L0–L3**) and qubits (**q0–q3**), shown on a shared color scale. (c,d) Corresponding parameter distributions using shared histogram bins. The encoder exhibits structured, layer-dependent parameter variation, indicating active and selective compression. The decoder parameters remain sharply concentrated near zero, revealing a pronounced asymmetry *consistent with compression-driven behavior*, reflecting strongly regularized and *minimal* reconstruction corrections.

**Fig. 7** reveals the following:

a) **Encoder parameters:** The encoder exhibits strong layer- and qubit-dependent structure, with pronounced heterogeneity emerging in deeper layers (notably L2). This indicates that compression is implemented through structured parameter organization rather than uniform rotations.

b) **Decoder parameters:** Decoder parameters remain uniform and close to zero across layers and qubits, reflecting strong regularization and a minimal corrective role during reconstruction.

c) **Encoder distribution:** A broad, asymmetric distribution with a heavy tail toward larger magnitudes confirms that the encoder actively explores parameter space to implement compression.

d) **Decoder distribution:** A sharply peaked distribution near zero confirms conservative reconstruction behaviour and reinforces the encoder–decoder asymmetry.

The encoder exhibits structured, layer- and qubit-dependent variation, with particularly strong heterogeneity emerging in the second encoder layer. In contrast, decoder parameters remain relatively uniform and concentrated near zero across all layers and qubits. The corresponding parameter distributions further highlight this difference, with the encoder displaying a broader, asymmetric distribution and the decoder exhibiting a sharply peaked distribution near zero. These observations indicate that compression is primarily implemented through structured encoder dynamics rather than decoder expressivity.

This asymmetry indicates that compression is primarily performed by the encoder, which actively redistributes information across qubits and layers, while the decoder applies minimal, targeted corrections during reconstruction. The corresponding parameter distributions further confirm that the encoder explores a significantly larger parameter space than the decoder, consistent with its role in information removal.

In the context of this specific Brain MRI dataset, the anomaly curve and the normal curve (**Fig. 11** of the Appendix), where they do not overlap, shows the tumorous features. A large overlap is not unexpected given the nature of the image data.

### 4.6 Visual inspection of QAE anomaly heatmaps

Beyond quantitative performance metrics, we examine the spatial structure of anomaly signals produced by the quantum autoencoder as shown in Fig. 8 which provides a qualitative illustration of the anomaly localization produced by the quantum autoencoder. The anomaly score clearly identifies the positions of the tumorous growths. Some further examples of visuals are shown in **Fig. 14** in Appendix

The left panel of **Fig. 8** shows an example DICOM slice which is raw brain MRI-like (no anomaly labels are used or required) serving as the reference anatomical context, while the right panel displays the corresponding QAE anomaly heatmap obtained by aggregating patch-level trash-based scores across the image. Regions with elevated anomaly scores are spatially localized and anatomically structured, indicating that anomalous signals arise from specific image regions rather than global intensity variations. This behavior is consistent with the compression-driven detection mechanism of the QAE, in which regions that cannot be efficiently compressed into the latent subspace yield higher anomaly scores.

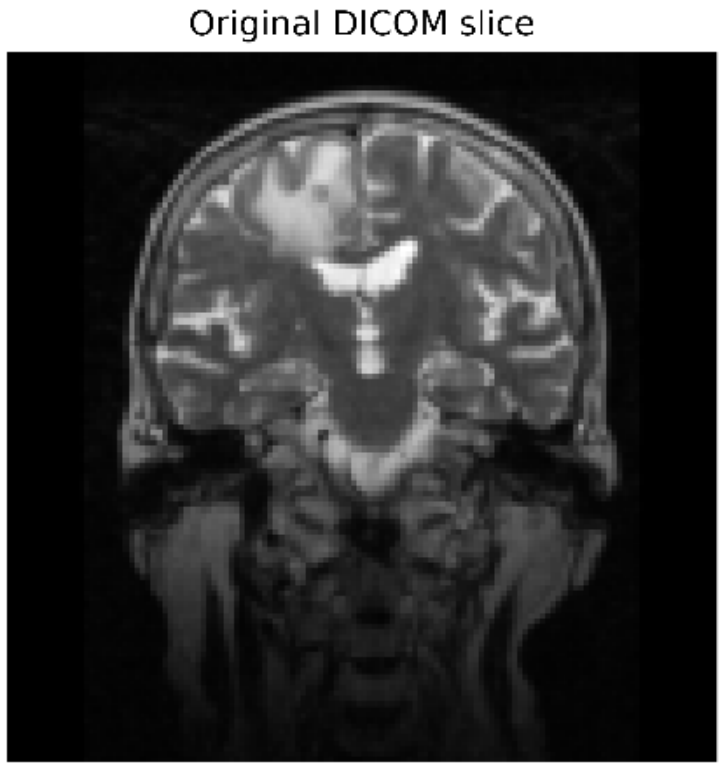

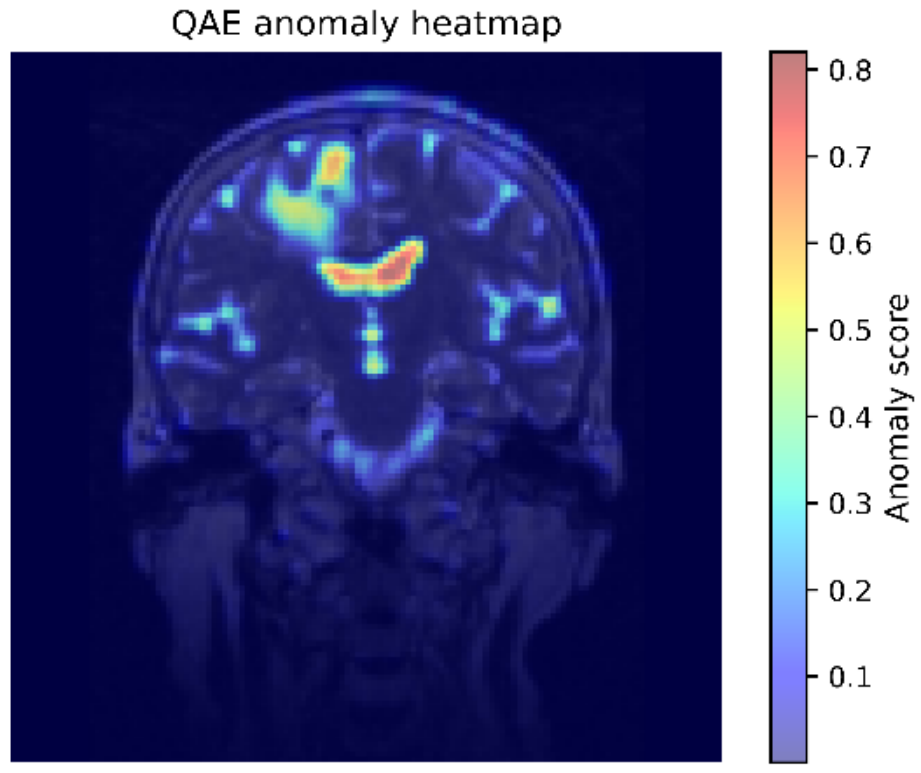

**Fig. 8.** Qualitative **anomaly localization using the quantum autoencoder.** *Left*: original DICOM slice. *Right*: anomaly heatmap generated from patch-level QAE anomaly scores, where warmer colors indicate higher incompressibility. The QAE produces spatially localized anomaly responses, highlighting specific anatomical regions rather than diffuse global deviations. Color overlay corresponds to the **QAE anomaly score.** Warmer colours (yellow–red) indicate **higher incompressibility,** while Cooler colours (blue) indicate **efficient compression**

The resulting heatmaps in **Fig. 8** demonstrate that the QAE produces interpretable spatial anomaly patterns that align with the underlying image structure. Encoder parameters exhibit pronounced layer- and qubit-dependent patterns, with structured activations and a broad distribution of values. In contrast, decoder parameters are sharply concentrated near zero, with only localized deviations. As indicated, the peak anomaly score in visual detection varies between ~0.5 and ~0.8 depending on the quality of the images, resolution and depth. High anomaly scores are *spatially localized*, not diffused. The QAE highlights specific anatomical regions rather than uniformly flagging the entire slice. This directly indicates that **anomaly detection arises from structured compression dynamics rather than reconstruction error**.

### 4.7 Per-qubit $L_2$ Magnitude Distribution

The per-qubit L2 magnitude distributions for normal and anomalous samples were obtained and were shown to exhibit consistent separation across all four qubits as shown in **Fig. 9**. Normal samples are concentrated at low magnitudes, whereas anomalous samples display broader, heavier-tailed distributions. This is reflected in nearly uniform Kolmogorov–Smirnov statistics (KS ≈ 0.36–0.37) and Wasserstein distances (W = 0.29) across qubits, indicating that the anomaly signal is distributed across the learned representation rather than localized to a single qubit.

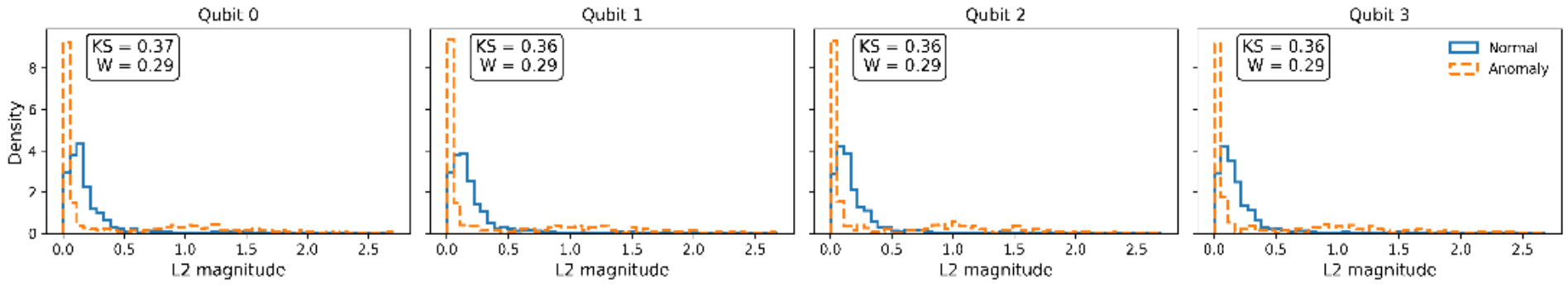

**Fig. 9.** Distribution of per-qubit $L_2$magnitudes for normal and anomalous samples across the four-qubit representation. Blue solid curves denote normal samples and orange dashed curves denote anomalous samples. For all qubits, normal samples remain concentrated near low $L_2$values, whereas anomalous samples exhibit broader, heavier-tailed distributions extending to larger magnitudes. The corresponding Kolmogorov–Smirnov statistics (KS ≈0.36–0.37) and Wasserstein distances ($W = 0.29$) are highly consistent across qubits, indicating that the anomaly-related deviation is distributed across the learned representation rather than localized to a single qubit. This representation-level separation is consistent with the spatial anomaly heatmaps, which show localized high-score regions in anomalous MRI slices.

The per-qubit $L_2$magnitude distributions provide additional evidence that the QAE learns a stable separation between normal and anomalous inputs at the representation level. As shown in **Fig. 9**, normal samples are concentrated at small $L_2$values, while anomalous samples exhibit broader and heavier-tailed distributions. The separation is quantitatively consistent across all four qubits, with Kolmogorov–Smirnov statistics of approximately $0.36 - 0.37$ and Wasserstein distances of $0.29$. This indicates that the anomaly signal is not dominated by a single qubit, but is instead distributed across the compressed latent representation.

### 4.8 Compression-driven anomaly detection in brain MRI

The proposed quantum autoencoder (QAE) yields a clear separation between normal and anomalous brain MRI slices when anomaly scores are aggregated at the slice level. While patch-level discrimination remains modest, as expected in a weak-label setting where many patches from anomalous slices are themselves normal, slice-level aggregation substantially improves separability. In particular, using calibrated patch scores together with top-$k$ aggregation produces a *slice-level* ROC–AUC of approximately **0.95**, indicating strong discrimination between normal and tumorous cases. This result contrasts with the patch-level behavior, where ROC–AUC remains between ~**0.813-0.54**, and highlights that the anomaly signal is concentrated in a limited subset of spatially informative patches rather than being uniformly distributed across the image.

The qualitative behavior of the model is consistent with this quantitative trend. As shown by the anomaly heatmaps, elevated scores are spatially localized and align with tumor-containing regions in anomalous MRI slices. These responses are not uniformly distributed over the image, nor are they driven solely by global intensity variation. Instead, the heatmaps exhibit anatomically structured hotspots, indicating that the QAE captures localized deviations relative to the learned normal manifold. This supports the interpretation that the slice-level improvement in ROC performance arises from the aggregation of sparse but high-confidence local anomaly responses.

Further insight is provided by the qubit-wise distribution analysis. The per-qubit $L_2$ magnitude distributions show a consistent separation between normal and anomalous samples across all four qubits. Normal samples remain concentrated near low $L_2$ values, whereas anomalous samples exhibit broader and heavier-tailed distributions. This separation is quantitatively reflected in nearly uniform Kolmogorov–Smirnov statistics (KS $\approx 0.36$–$0.37$) and Wasserstein distances ($W = 0.29$) across qubits. The consistency of these values suggests that anomaly-related information is not localized to a single qubit, but distributed across the compressed representation.

Taken together, the ROC analysis, spatial heatmaps, and qubit-level statistics provide a coherent picture of the learned anomaly mechanism. Localized tumor-related deviations in the image domain produce sparse high-score regions in the heatmaps, and these in turn correspond to a stable distributional shift in the latent quantum representation. This agreement between image-space localization and qubit-space statistics supports the interpretation that the QAE detects anomalies *through structured incompressibility relative to normal data*, rather than through reconstruction error alone.

### 4.9 General Brain MRI Data Behaviour

The anomaly score varied between 0.002 for the class SE000004 and 0.405 for class SE000007. The distribution shows a close relationship between the normal data and the data with tumorous anomaly. This is not unexpected given the greyscale Brain MRI images are visually generally similar in nature, the span of the tumorous growths much smaller in comparison to the actual resolution of the images; although we did normalise the data to try and minimise the effect of the disparity in sizes, the contextual similarities cannot be ignored – and the wide variance in the distribution of resolution in the input image dataset is an additional factor to be considered for the anomaly scores. A recorded Youden's J threshold (**Fig. 11**-**Fig. 13**) between 0.002 and 0.0034 with values of J varying between 0.1733 for classes SE000001 and 0.4462 for class SE000002 shows that the discriminative power of the classification is low (except for class SE000007 with a threshold of 0.405) as shown in **Fig. 7**. However, the Youden J [136] index is a simple and widely used method, its implicit assumption of equal misclassification costs and its relation to prevalence may not be suitable for all clinical or machine learning contexts. It is also to be kept in mind that the "normal" dataset and the tumor/anomaly dataset in this case, were obtained from different patients, recorded by different doctors on different MRI equipment at different times with differing resolutions. In the context of this particular Brain MRI dataset, the anomaly curve and the normal curve, where they do not overlap, shows the tumorous features. A large overlap is not unexpected given the nature of the image data. The anomaly plots for QAE, classical autoencoder and PCA comparisons are shown in **Fig. 11-Fig. 13** in the Appendix.

Finally, the qubit-wise $L_2$ distributions in **Fig. 9** and the spatial anomaly heatmaps in **Fig. 8** should be interpreted as complementary views of the same effect. The former shows that anomalous inputs induce a consistent distributional shift across all qubits, whereas the latter shows where these deviations occur in the image domain. Taken together, these results suggest that localized tumor-related abnormalities are encoded as structured, distributed deviations in the QAE latent representation.

### 4.10 Discussions

This work investigates whether a quantum autoencoder (QAE) can serve as a clinically relevant assistive tool for anomaly detection in brain MRI data. To our knowledge, this represents one of the first applications of quantum machine learning methods to anomaly detection in labeled brain MRI images. The resulting anomaly heatmaps demonstrate that the QAE is capable of identifying tumor regions when compared against normal reference images, producing spatially localized responses that align with clinically relevant structures.

**Compression-Driven Anomaly Detection.** This study investigates whether a quantum autoencoder (QAE) can provide a principled mechanism for unsupervised anomaly detection in brain MRI data. As such, the focus is on understanding how **quantum compression dynamics** can be exploited to detect localized deviations in complex medical images. Given that tumorous regions occupy a small fraction of the image relative to the skull and surrounding anatomy, effective detection requires sensitivity to local structural deviations rather than global image statistics.

**Role of Encoder Structure and Parameter Organization.** Our results show that effective anomaly detection in the QAE emerges from **structured allocation of parameter capacity within the encoder**, rather than from increased decoder expressivity or reconstruction accuracy alone. Parameter heatmaps and layer-wise energy analysis reveal a consistent encoder–decoder asymmetry, where information compression is implemented through specific encoder layers. Importantly, the total normalized parameter energy remains comparable between encoder and decoder components, indicating that anomaly detection is governed by *where* and *how* parameters are deployed, rather than by overall model capacity.

**Per qubit L2 Magnitude Distributions.** The consistency of the KS (Kolmogorov-Smirnov**)** and W (Wasserstein) statistics across all qubits suggests that anomaly detection in the QAE is not driven by an isolated latent component, but by a distributed reorganization of the compressed representation. This is important for **interpretability**: the model does not rely on a fragile single-qubit response, but instead encodes anomalous structure through a stable multi-qubit deviation pattern. The spatial heatmaps support this interpretation by showing that the corresponding anomaly signal remains anatomically localized in the image space, particularly around tumorous regions. In this sense, the qubit-level statistics and the heatmaps provide complementary evidence for compression-driven anomaly detection.

**Comparison with Classical Baselines.** In contrast to classical autoencoder and PCA baselines, which are driven by reconstruction fidelity or linear variance truncation, the QAE produces anomaly scores grounded in **incompressibility with respect to a learned latent representation**. This distinction is evident both quantitatively and qualitatively. The spatial anomaly heatmaps generated by the QAE exhibit localized, anatomically structured responses, indicating that detection is driven by region-specific deviations rather than trivial global intensity differences.

This behavior highlights a key distinction between quantum and classical autoencoder-based anomaly detection. Rather than relying on reconstruction error alone, the QAE explicitly learns to compress normal data by driving designated trash qubits toward a fixed reference state. Inputs that resist this compression naturally yield elevated anomaly scores, producing a controlled and interpretable compression–reconstruction trade-off characterized by a clear knee operating point. This property enables principled threshold selection for anomaly detection.

**Practical Implications.** From a practical perspective, the visualization of anomaly scores overlaid on DICOM slices demonstrates clear localization of tumorous regions. While the proposed approach is not intended to replace clinical diagnosis, it has the potential to function as a *decision-support tool* that enhances speed, consistency, and triage efficiency in high-volume clinical environments. Final diagnostic decisions remain dependent on expert interpretation, clinical context, and histopathological confirmation.

**Limitations and outlook.** All experiments were conducted using a noiseless state-vector simulator and a small number of qubits, and the analysis was restricted to patch-based representations. Accordingly, the results should be interpreted as a conceptual demonstration of compression-driven anomaly detection. Future work will address scalability, robustness under realistic noise models, and evaluation on quantum hardware.

## 5 Conclusion

In this work, we investigated a quantum autoencoder (QAE) as a **compression-driven framework for unsupervised anomaly detection** in brain MRI data. By explicitly training the model to discard information via auxiliary trash qubits, the QAE exhibits a controlled compression–reconstruction trade-off characterized by a clear knee operating point. Analysis of the learned parameters reveals a pronounced encoder–decoder asymmetry, with compression implemented through structured allocation of parameter capacity within the encoder rather than through increased parameter magnitude or decoder expressivity. This structure provides a mechanistic explanation for both the stability of the anomaly detection signal and the emergence of a principled operating regime.

We demonstrated the approach on publicly available, high-resolution brain MRI DICOM datasets and obtained competitive slice-level performance, with composite AUC and accuracy values exceeding those of classical autoencoder and PCA baselines. Unlike classical methods, whose anomaly scores exhibit heavy overlap between normal and anomalous samples, the QAE produces well-separated score distributions and spatially localized anomaly heatmaps. These results indicate that anomaly detection in the QAE is driven by **incompressibility with respect to a learned latent representation**, rather than by reconstruction error alone.

Consistent qubit-wise distribution shifts, together with spatially localized anomaly heatmaps, indicate that the proposed QAE captures tumor-related deviations as structured changes in the compressed latent representation rather than as isolated or purely reconstruction-driven effects.

The results highlight how quantum autoencoders provide an interpretable and controllable framework for studying compression dynamics in quantum machine learning. The explicit separation between compression and reconstruction, together with circuit-level parameter analysis, *offers insights that are less accessible* in classical architectures.

From a practical perspective, the localized anomaly heatmaps suggest potential utility as a decision-support tool for assisting clinicians in triage and screening workflows. Final diagnosis remains dependent on expert interpretation and clinical context. Overall, this study establishes quantum autoencoders as a principled platform for exploring compression-driven anomaly detection and motivates further investigation under realistic noise models and on quantum hardware.

Future work will focus on scaling to larger circuits, incorporating realistic noise models, and evaluating robustness on quantum hardware. Extending the approach beyond patch-based representations toward end-to-end image-level quantum models, as well as exploring hybrid quantum–classical workflows in clinical settings, represent important directions for further investigation.

### 5.1 Summary and Outlook

QAE structures studied here gives rise to stable anomaly detection performance and depending on data types, spatially interpretable anomaly heatmaps. While no quantum advantage is claimed, these results demonstrate that quantum autoencoders offer a transparent and controllable mechanism for anomaly detection, making them a promising, performing tool for studying information compression in quantum machine learning.

Future work will focus on scaling to larger circuits, incorporating realistic noise models, and evaluating robustness on quantum hardware. Extending the approach beyond patch-based representations toward end-to-end image-level quantum models, as well as exploring hybrid quantum–classical workflows in clinical settings, represent important directions for further investigation.

## 6 Acknowledgement

S. Ganguly is grateful to *Dr. Saumyadip Dasgupta, MD*. [130], Carmel, Indiana, United States, for his valuable feedback on the quality of the DICOM data anomaly detection and its usability in medicine.

# 7 Appendix

## 7.1 Details of experimental setup

**Datasets:** Experiments were conducted on publicly available brain MRI DICOM datasets obtained from the Hugging Face repository cited in Ref. [113]. The datasets contain both normal and tumorous grayscale brain MRI slices acquired under heterogeneous conditions, including variation in patient population, scanner type, acquisition time, and image resolution. This heterogeneity increases the practical relevance of the anomaly detection task, since normal and anomalous samples are not matched by acquisition protocol alone. For the anomaly detection setting considered here, slices labeled as normal were used to define the reference distribution, while slices containing tumorous regions were treated as anomalous during evaluation. All labels were inherited from the original medical annotations downloaded with the datasets.

**Preprocessing and patch extraction:** Each DICOM slice was converted to a single-channel floating-point image and normalized on a per-slice basis. Images were resized to $128 \times 128$ pixels. Each slice was then decomposed into overlapping $2 \times 2$ patches. For training, a patch stride of **1** was used, whereas for slice-level evaluation a stride of **4** was used to reduce computational cost. Each patch was flattened into a **4-dimensional** feature vector and rescaled to the interval $[0, \pi]$, enabling direct angle encoding into a four-qubit quantum circuit. Because tumorous regions occupy only a localized subset of the full MRI slice, anomaly detection was formulated at the patch level and subsequently aggregated to the slice level.

**Quantum autoencoder configuration:** The model is trained using a trash-qubit loss that enforces compression by driving the auxiliary qubits to the $|00\rangle$ state. The QAE consists of four qubits arranged into two logical qubits and two auxiliary (trash) qubits. Classical input features are encoded using single-qubit rotation gates. The encoder comprises of two layers of parameterized single-qubit rotations followed by nearest-neighbor entangling gates, while the decoder mirrors this structure to reconstruct the input state. Measurements are performed in the computational basis to obtain output probabilities. Reconstruction performance is evaluated separately using a probability-based reconstruction metric.

**Training**: Training was carried out in a *hybrid quantum–classical workflow* using PennyLane [57] and PyTorch on a noiseless state-vector simulator. Optimization used the Adam optimizer with learning rate $10^{-2}$ batch size 32, and 10 epochs. The model was trained using normal patches only.

**Training objective:** The QAE was optimized to maximize the compressibility of normal inputs by driving the trash qubit toward a fixed reference state. To discourage trivial compression, a pseudo-anomaly term based on *feature dropout corruption* was introduced. The corruption probability was set to 0.5, and the pseudo-anomaly loss was weighted by $\alpha = 0.2$. This yielded a contrastive objective that encouraged strong compression of normal data while separating corrupted inputs from the learned normal manifold.

**Patch-level scoring and slice-level aggregation:** Patch-level scores were derived from the behavior of the trash qubit. Slice-level anomaly scores were obtained using **top-$k$** aggregation with $k = 0.05$, i.e. the highest-scoring **5%** of patches within each slice were averaged. This choice reflects the spatial sparsity of tumorous regions. Before aggregation, patch scores were calibrated using a **power temperature transform**, $s_\tau = s^\tau$, with candidate values

$$\tau \in \{0.2, 0.3, 0.5, 0.7, 1.0\}.$$

The best-performing temperature was selected on a validation subset. Best $\tau$ used for the reported result**:** 0.7.

**Baselines:** Two classical baselines were considered under the same preprocessing and slice aggregation pipeline: **principal component analysis (PCA)** and a **classical autoencoder**. Both baselines operated on the same preprocessed patch representation and were trained using normal data only. This ensured that performance differences reflected the modeling framework rather than differences in input features or evaluation procedure.

**Evaluation protocol:** Performance was evaluated at both the patch and slice levels. Patch-level discrimination was summarized using ROC–AUC, but was interpreted cautiously because many patches extracted from anomalous slices remain visually normal. Slice-level discrimination was treated as the primary endpoint and assessed using **ROC curves**, **AUC**, and threshold-based accuracy.

To generate spatial anomaly maps, patch-level scores were reprojected onto their original image locations to form **slice-level heatmaps**. These were used for qualitative assessment of whether the learned anomaly signal aligned with tumorous regions.

**Statistical analysis:** Latent-space differences between normal and anomalous samples were analyzed using qubit-wise $L_2$ magnitude distributions. Distributional separation was quantified using the **Kolmogorov–Smirnov statistic** and the **Wasserstein distance**. Slice-level statistical significance was assessed using nonparametric permutation testing, and confidence intervals for slice-level AUC were estimated using bootstrap resampling over slices.

**Software and hardware:** All experiments were implemented in Python 3.11.9 using PennyLane 0.43.1, PyTorch 2.9.1, NumPy 2.3.5, SciPy 1.16.3, scikit-learn 1.8.0,

Matplotlib 3.10.7, scikit-image 0.25.2, and pydicom 3.0.1. The software environment was Ubuntu 24.04 LTS with CUDA.

### 7.2 Slice-ROC Result

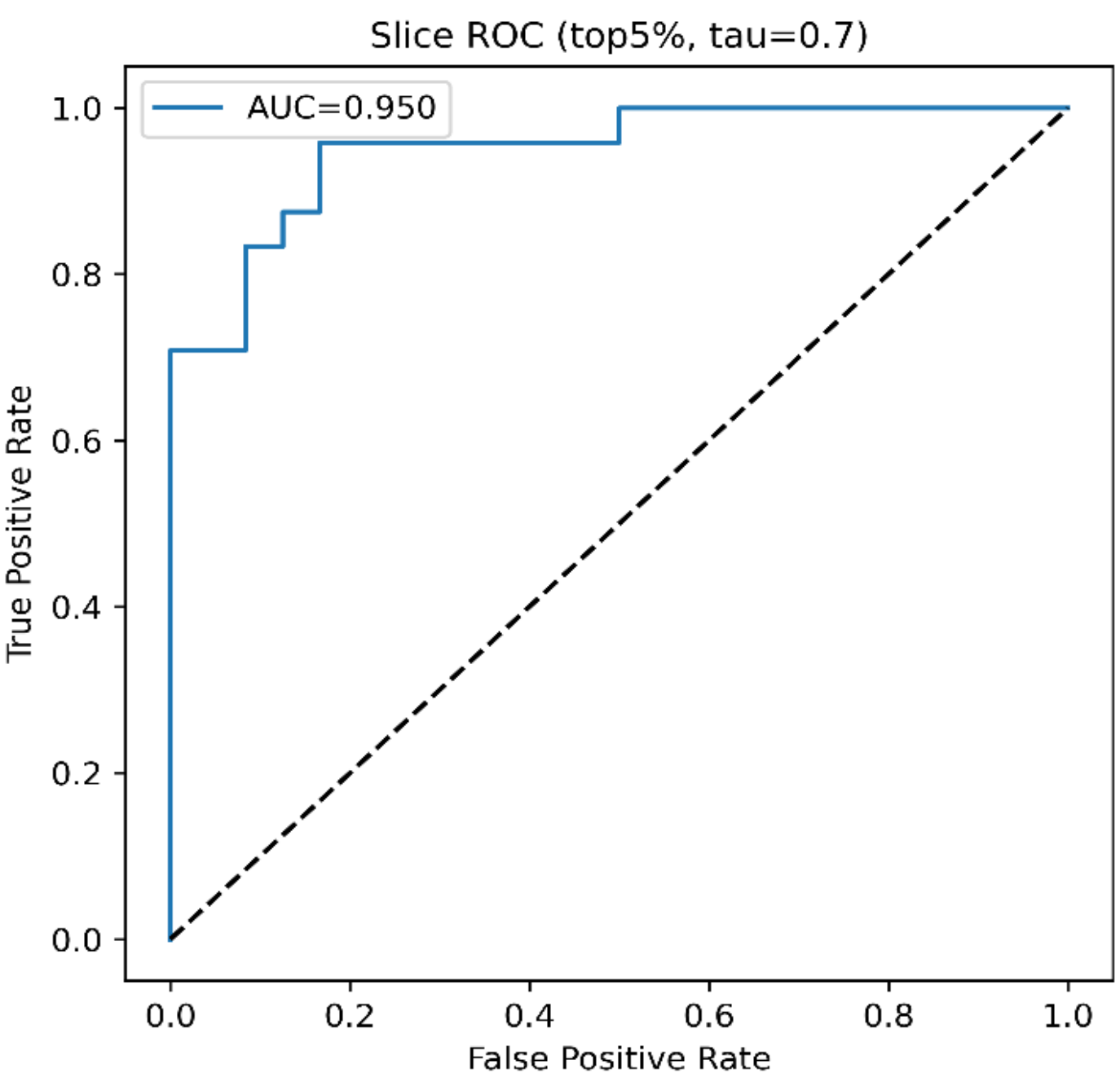

**Fig. 10.** Slice-level ROC curve for the QAE using top-5% patch aggregation and $\tau = 0.7$. The model achieves **AUC = 0.950**, demonstrating excellent discrimination between normal and anomalous MRI slices.

This ROC (ref. Section 3.6) curve indicates excellent slice-level discrimination between normal and anomalous MRI slices. The reported AUC = 0.950 shows that the aggregated QAE score separates the two classes very effectively. Several features are notable:

- The curve rises **very steeply near the origin**, which means the model achieves a high true-positive rate at low false-positive rates.
- From the plot, the true-positive rate is already around **0.7** at essentially **zero false positives**, and exceeds **0.8–0.9** at relatively small false-positive rates.
- The curve reaches **TPR ≈ 1.0** well before the maximum false-positive rate, indicating that nearly all anomalous slices can be recovered with an appropriate threshold.
- The **step-like shape** suggests a limited number of evaluation slices, which is typical for slice-level ROC analysis.

In terms of interpretation, this result supports the claim that **top-5% aggregation combined with temperature calibration ($\tau = 0.7$)** successfully amplifies sparse but informative patch-level anomaly responses into a strong slice-level decision signal. This is consistent with the heatmap results, where only a limited subset of patches contributes strongly to the anomalous slice score.

**Description of $\boldsymbol{\tau}$.** Here, $\tau$ is the **temperature-scaling parameter** used to recalibrate patch scores before slice-level aggregation, used in a transform like:

$$s_\tau = s^\tau$$

where $s$ is the patch score, in our case typically $p_{\text{good}}$ (refer Section 3.2). $\tau$ is *not a trainable quantum parameter*. It is a *post-processing hyperparameter* chosen after training, usually by testing several values on a validation/tuning set and selecting the one that gives the best slice-level AUC.

**Interpretation.** Using **top-5% aggregation** of calibrated patch scores ( $\tau = 0.7$ ), the proposed QAE achieved a **slice-level ROC–AUC of 0.950**, demonstrating strong discrimination between normal and anomalous brain MRI slices. This result is consistent with the spatial anomaly heatmaps, which show that tumor-related signal is **localized to a relatively small subset of high-scoring patches** rather than distributed uniformly across the slice. Top-$k$ aggregation is therefore well matched to the imaging problem: because tumorous regions occupy only a limited portion of the full MRI slice, averaging over all patches would dilute the anomaly signal, whereas selecting the highest-scoring fraction preserves the most informative local responses.

### 7.3 Anomaly Plots: QAE, Classical & PCA

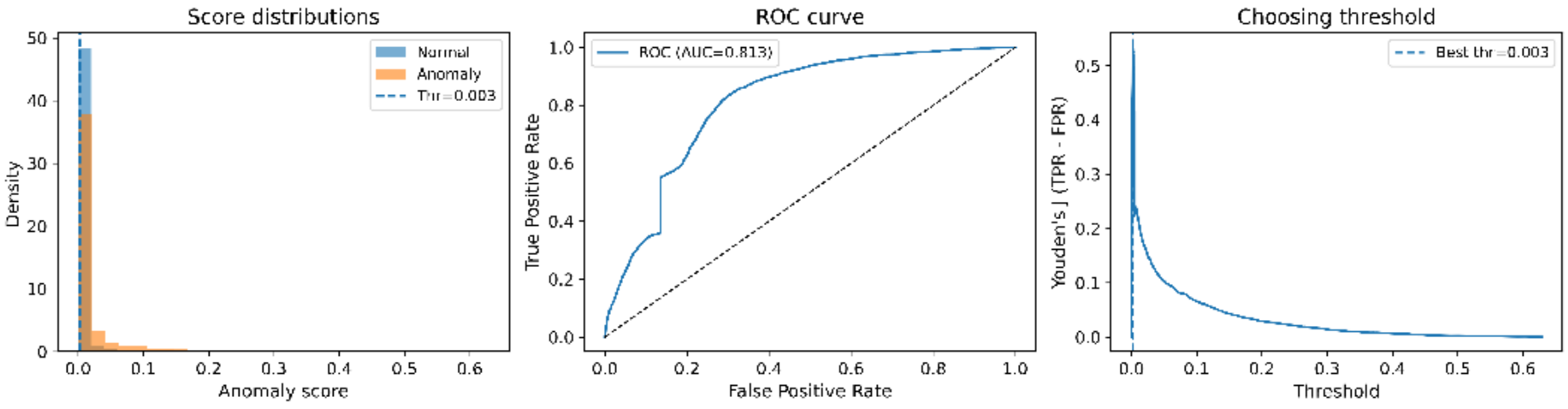

**Fig. 11.** QAE Anomaly Score Plots

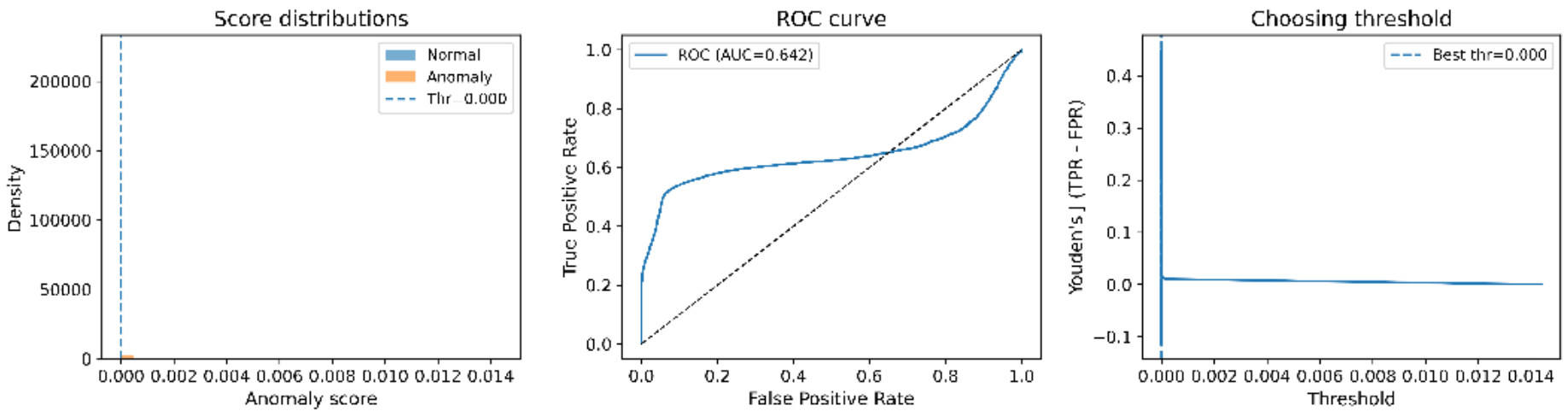

**Fig. 12.** Classical Anomaly Plot

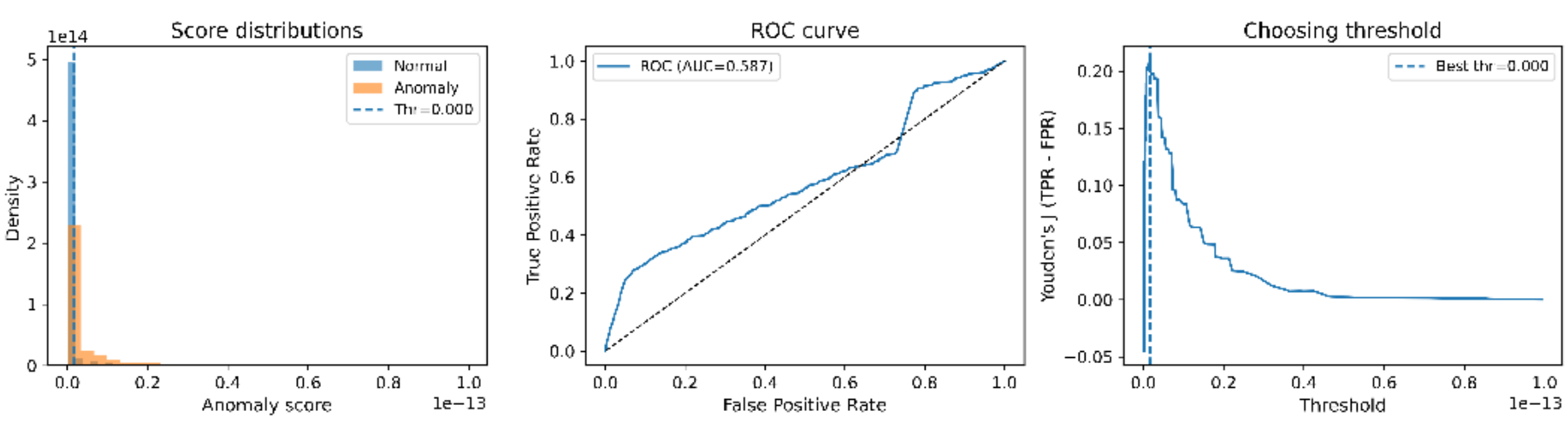

**Fig. 13.** PCA Anomaly Plot

## 7.4 Anomaly Heat Detection: Further Results

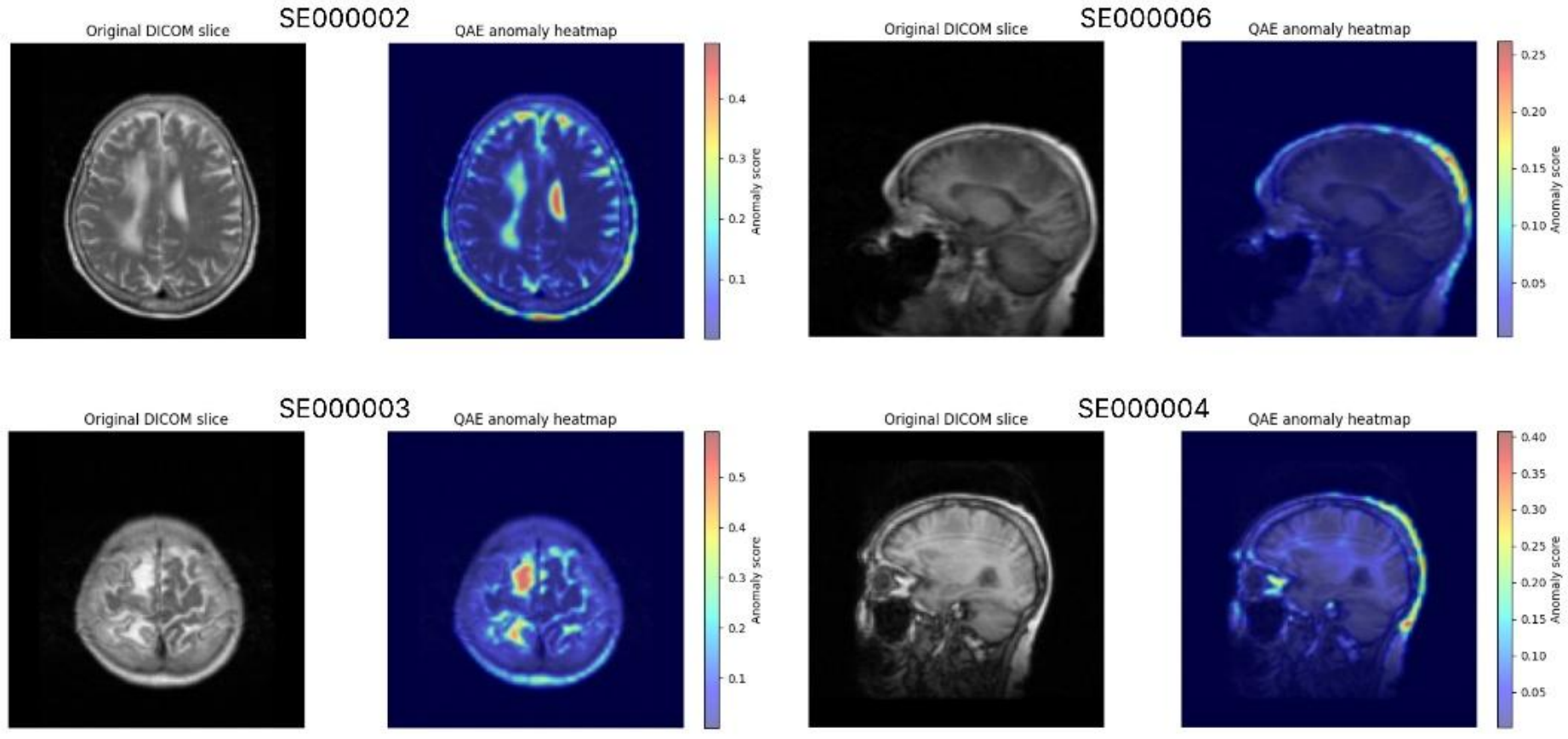

**Fig. 14.** Four examples of visual anomaly detection in Brain MRI plates from four classes of the DICOM data. The classes are marked on top for each pair: SE000002, SE000003, SE000004 and SE000006. To be noted is the similarity in visual structure between the pair of original plates of SE000002 and SE000003, and the pair of SE000004 and SE000006. Our QAE successfully detects the tumors as anomalies between the pairs at different areas of the corresponding different brains and successfully indicates the possible existence of the disease as a time-saving tool for physicians. QAE based enhancement facilitates the accuracy of the anomaly detection by helping to stabilize the model performance via uniform distribution of data within the encoder and the decoder, both of which are quantum.

### 7.5 QAE model pipeline

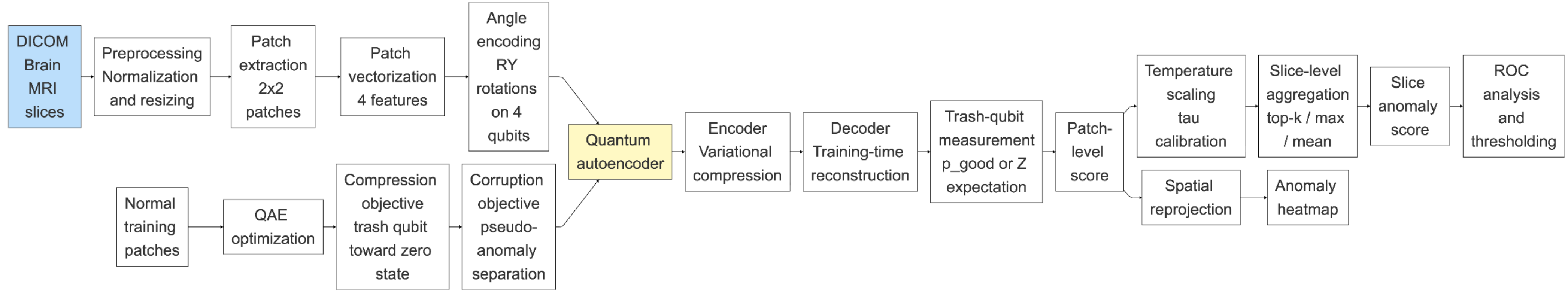

**Fig. 15.** End-to-end pipeline of the proposed quantum autoencoder–based anomaly detection framework. Brain MRI DICOM slices are preprocessed and decomposed into small image patches, which are encoded into a 4-qubit quantum state via angle encoding. A variational quantum autoencoder produces patch-level anomaly-related scores through trash-qubit behavior, which are calibrated and aggregated to obtain slice-level anomaly scores. Patch scores are also reprojected to generate spatial anomaly heatmaps. Produced using Mermaid [138]